\documentclass[floatfix,nofootinbib]{revtex4-2}

\usepackage{graphicx}
\usepackage{dcolumn}
\usepackage{bm}

\usepackage{color}
\usepackage{amsmath}
\usepackage{amsfonts}
\usepackage{amssymb}
\usepackage{latexsym}
\usepackage{caption}
\usepackage{subcaption}
\usepackage{tikz}
\usepackage{circuitikz}
\usepackage[compat=1.1.0]{tikz-feynman}
\usepackage[compat=1.1.0]{tikz-feynhand}
\usepackage{hyperref}
\definecolor{darkred}{rgb}{0.8,0.1,0.1}
\hypersetup{colorlinks=true, linkcolor=darkred, citecolor=blue, linktoc=page}
\usepackage{tikz-cd,mathtools}

\def\s{\sigma}
\def\m{\mu}
\def\n{\nu}
\def\l{\lambda}
\def\){\right)}
\def\({\left( }
\def\]{\right] }
\def\[{\left[ }

\newcommand{\be}{\begin{equation}}
\newcommand{\ee}{\end{equation}}

\def\bsb{{\boldsymbol{b}}}
\def\bsk{{\boldsymbol{k}}}
\def\bsp{{\boldsymbol{p}}}
\def\bsq{{\boldsymbol{q}}}

\def\bsx{{\boldsymbol{x}}}
\def\bsy{{\boldsymbol{y}}}

\def\bsX{{\boldsymbol{X}}}

\def\no{\nonumber}
\def\p{\partial}

\def\ep{\varepsilon}

\tikzset{graviton/.style={decorate, decoration={snake, amplitude=.4mm, segment length=1.5mm, pre length=.5mm, post length=.5mm}, double}}

\begin{document}

\title{Universal features of \texorpdfstring{$ 2\to N$}{} scattering in QCD and gravity from shockwave collisions}

\author{Himanshu Raj}
\email{himanshu.raj@stonybrook.edu}
\affiliation{
Universit\'e Paris-Saclay, CNRS, CEA, Institut de Physique Th\'eorique, 91191, Gif-sur-Yvette, France\\
Center for Frontiers in Nuclear Science, Department of Physics and Astronomy, Stony Brook University, Stony Brook, NY 11794, USA
}

\author{Raju Venugopalan}
 \email{raju.venugopalan@gmail.com}
\affiliation{
Department of Physics, Brookhaven National Laboratory, Upton, NY 11973, USA\\
Center for Frontiers in Nuclear Science, Department of Physics and Astronomy, Stony Brook University, Stony Brook, NY 11794, USA}

\date{\today}

\begin{abstract}
A remarkable double copy relation of Einstein gravity to QCD in Regge asymptotics is $\Gamma^{\mu\nu}= \frac12C^\mu C^\nu- \frac12N^\mu N^\nu$, where $\Gamma^{\mu\nu}$ is the gravitational Lipatov vertex in the $2\to 3$ graviton scattering amplitude, $C^\mu$ its Yang-Mills  counterpart, and $N^\mu$ the QED bremssstrahlung vertex. In QCD, the Lipatov vertex is a fundamental building block of the BFKL equation describing $2\to N$ scattering of gluons at high energies. 
Likewise, the gravitational Lipatov vertex is a key ingredient in a 2-D effective field theory framework describing trans-Planckian $2\to N$ graviton scattering. 
We construct a quantitative correspondence between a semi-classical Yang-Mills framework for  radiation in gluon shockwave collisions and its counterpart in general relativity. In particular, we demonstrate the Lipatov double copy in a dilute-dilute approximation corresponding to $R_{S,L}$, $R_{S,H}$ $ \ll b$, with $R_{S,L}$, $R_{S,H}$ the respective emergent Schwarzchild radii generated in shockwave collisions and $b$ is the impact parameter. We outline extensions of the correspondence developed here to the dilute-dense computation of gravitational wave radiation in close vicinity of one of the black holes, the construction of graviton propagators in the shockwave background, and a renormalization group approach to compute $2\rightarrow N$ amplitudes that incorporates graviton reggeization and coherent graviton multiple scattering. 
\end{abstract}

\maketitle


\section{Introduction}
\label{sec:intro}

The discovery of the AdS/CFT correspondence between $\mathcal{N}=4$ supersymmetric Yang-Mills theory and string theory in $AdS_5\times S^5$ spacetime triggered great interest in quantitative relations between QCD-like theories and gravity in a variety of geometries \cite{Maldacena:1997re}. A significant subsequent development\footnote{A brief general introduction can be found in \cite{White:2017mwc}.} is the quantitative color-kinematics duality between perturbative QCD amplitudes and amplitudes in Einstein gravity discovered by Bern, Carrasco and Johansson (BCJ) \cite{Bern:2008qj}.  The origins of this duality with possible extensions to loop amplitudes in supersymmetric variants of these theories \cite{Bern:2010ue, Bern:2019prr}, can be traced to the work of Kawai, Lewellyn and Tye (KLT) who found that tree-level n-point closed string amplitudes can be written as sums over products of open string amplitudes\footnote{This was recently extended to one-loop string amplitudes in \cite{Stieberger:2023nol}.}. At energies much below the inverse string scale $\lambda_s$ where string theory reduces to quantum field theory, this double copy relates a tree level four point amplitude in gravity to tree level four point amplitudes in Yang-Mills theory~\cite{Kawai:1985xq}. Similar KLT-type relations derived for higher point string amplitudes can be generalized in the field theory limit to arbitrary numbers of external particles. For discussions of the extension of these dualities to higher loop orders, we refer the reader to \cite{Bern:2019prr}.

Though not as widely known, a double copy between gravitational amplitudes and QCD amplitudes in the Regge asymptotics of both theories was derived previously by Lipatov in two remarkable papers prior to the KLT work~\cite{Lipatov:1982it,Lipatov:1982vv}. Specifically as we shall discuss at length, Lipatov observed that the effective gravitational vertex that represents the sum of all $2\rightarrow 3$ amplitudes can be expressed as a difference of two terms, one representing the bilinear of its Yang-Mills $2\rightarrow 3$ counterpart and the other the bilinear of the QED bremsstrahlung vertex. In QCD this effective vertex widely known as the Lipatov vertex is a key building block of the $2\rightarrow N$ scattering amplitude in Regge asymptotics $s\gg |t|$, where $s$ is the squared center-of-mass energy and $-t$ is the squared momentum transfer. The Lipatov vertex together with the``reggeization" of the t-channel gluon propagator generates the celebrated BFKL equation~\cite{Kuraev:1977fs,Balitsky:1978ic} for the perturbative pomeron 
describing the energy evolution of high energy cross-sections to leading logarithmic (LLx) and next-to-leading logarithmic (NLLx) accuracy in Bjorken $x$, for $x\rightarrow 0$ in the Regge limit
\cite{DelDuca:1995hf,DelDuca:2022skz}. For the interpretation of the Lipatov double copy in terms of the BCJ color-kinematics duality, we refer the reader to \cite{SabioVera:2012zky,Johansson:2013nsa,SabioVera:2014mkb}.

The common features of Lipatov vertices and reggeized propagators formed the basis of Lipatov's 2-D reggeon effective field theory (EFT) for QCD and gravity~\cite{Lipatov:1991nf}. Based on this work, and their previous work in this direction~\cite{Amati:1987uf,Amati:1990xe,Amati:1992zb}, Amati, Ciafaloni and Veneziano (ACV)~\cite{Amati:1993tb,Amati:2007ak} significantly developed the 2-D EFT to compute 
the S-matrix for eikonal scattering and inelastic graviton emission as a function of impact parameter in the kinematic region $b > R_S > \lambda_s$, where $b$ is the impact parameter of the scattering, $R_S$ the Schwarzchild radius, and $\lambda_s$ the string length scale. An interesting question in this regime is whether gravitational collapse due to the overoccupancy of soft gravitons produced occurs at a critical impact parameter $b_c\sim R_S$ leading to black hole formation. For some of the subsequent literature in this direction, see for instance \cite{Addazi:2016ksu,Ciafaloni:2017ort,Ciafaloni:2016nul,Ciafaloni:2018uwe} and references within. 

In examining and developing the chain of ideas on $2\rightarrow N$ trans-Planckian scattering and black hole formation in gravity (which can be traced to the seminal S-matrix investigations of black holes in \cite{tHooft:1984kcu,Dray:1984ha,tHooft:1987vrq,tHooft:1996rdg,Gross:1987pd}) 
we are motivated both by the aforementioned quantitative double copy connections as well as highly suggestive qualitative, and seemingly universal, features of high energy scattering in QCD and gravity that emerge when the occupancy $N\gg 1$ that were discussed in \cite{Dvali:2021ooc}. On the gravity side of this semi-classical correspondence, it was argued that $2\rightarrow N$ trans-Planckian scattering leads to the formation of black holes understood as semi-classical lumps of size $R_S$ that saturate unitarity when 
$\alpha \,N\sim 1$, where $\alpha=Q^2/M_{\rm Pl}^2$ for a probe with momentum resolution $Q\sim 1/R_S$~\cite{Dvali:2011aa,Dvali:2014ila}. (Note $M_{\rm Pl}^2 = \hbar/G$, where $G$ is Newton's constant.) Such black hole semi-classical lumps saturate S-matrix unitarity if and only if the microstate entropy in the scattering is $S=1/\alpha$~\cite{Dvali:2020wqi}. This entropy saturates the Bekenstein-Hawking area law where $M_{\rm Pl}$ is interpreted as the Goldstone scale corresponding to the breaking of Poincar\'{e} invariance~\cite{Dvali:2015ywa}. The connections between the semi-classical ``Quantum N-Portrait" (BHNP) of black hole dynamics and the ACV approach have been discussed previously in \cite{Dvali:2014ila,Addazi:2016ksu}.

The QCD side of the semi-classical correspondence can be traced back forty years to the 
observation that in the Regge asymptotics of $s\gg |t| \gg \Lambda_{\rm QCD}^2$, gluon distributions in hadrons saturate at a maximal occupancy of $N=1/\alpha_S$ in a region of screened color charge of size $1/Q_S$, where $Q_S\gg \Lambda_{\rm QCD}$ is the emergent saturation scale~\cite{Gribov:1983ivg,Mueller:1985wy}. The many-body dynamics of gluon saturation is described in the Color Glass Condensate (CGC) EFT~\cite{Gelis:2010nm,Iancu:2003xm,Kovchegov:2012mbw} where a large number of fast (light cone) color sources (as for example in a nucleus with atomic number $A\gg 1$) source high occupancy gauge fields, with the latter existing on parametrically much shorter time scales than the former; the explicit construction of the EFT for large nuclei was performed in \cite{McLerran:1993ka,McLerran:1993ni,McLerran:1994vd,Jeon:2004rk}. The kinematic separation of fast and slow modes in the EFT naturally leads to a Wilsonian renormalization group (RG) framework describing the evolution in rapidity of the separation between color sources and fields~\cite{Balitsky:1995ub,Jalilian-Marian:1997qno,Jalilian-Marian:1997ubg,Jalilian-Marian:1998tzv,Iancu:2000hn,Iancu:2001ad,Ferreiro:2001qy}. 

In the infinite momentum frame, the semi-classical lump as viewed by a probe is a gluon shockwave and scattering cross-sections in this regime can be constructed from n-point Wilson line correlators in the shockwave background. The nonlinear RG equations describing their evolution with rapidity capture the physics of eikonalized multiple scattering contributions as well as inelastic gluon emission treated on the same footing\footnote{The RG picture in the CGC EFT can be mapped to the 2-D reggeon EFT developed by Lipatov~\cite{Lipatov:1996ts} that we discussed earlier; see for instance \cite{Hentschinski:2020rfx}. }. For the simplest two-point correlator describing the inclusive deeply inelastic scattering (DIS) cross-section at fixed impact parameter, the corresponding RG equation has a closed form nonlinear form (for $A\gg 1$ and $N_c\gg 1$) called the Balitsky-Kovchegov (BK) equation~\cite{Balitsky:1995ub,Kovchegov:1999yj}. The BK equation has a nontrivial fixed point that unitarizes the cross-section at maximal occupancy~\cite{Kovchegov:1999ua,Iancu:2003xm}; its behavior is described by the energy-dependent saturation scale $Q_S$. In the weak field limit where nonlinear corrections are small, the BK equation reduces to the BFKL equation we discussed previously. 

The takeaway message from this discussion is that the semi-classical CGC EFT quantitatively reproduces results from the perturbative QCD Feynman diagram results for $2\rightarrow N$ scattering. Further, it  enables one (in the high occupancy regime of large number of sources, or $A\gg 1$) to go well beyond by including the physics of {\it both} inelastic radiation and multi-Pomeron interactions, the latter being essential for unitarization of the cross-section. Computations in the CGC EFT including all-order power corrections are now at NLLx accuracy and provide a quantitative description of multiparticle production at collider energies~\cite{Morreale:2021pnn}. 

The question we shall address beginning with this work and in subsequent work is whether the framework of semi-classical strong field EFTs can analogously be applied to gravity in Regge asymptotics. This approach is complementary to that of ACV and can be regarded as a  quantitative realization of the ideas discussed in the context of gravity in \cite{Dvali:2011aa,Dvali:2014ila,Dvali:2015ywa,Dvali:2020wqi} and the connections to the CGC EFT noted in \cite{Dvali:2021ooc}. We emphasize that while double copies - in particular the classical double copy we shall discuss shortly - provide a powerful guide in this investigation,  the development of the desired semi-classsical EFT is not necessarily limited to the regime of their applicability. This is because  ``classical" in our context refers to the regime of very high occupancies ($N\gg 1$) - or strong fields.  

\begin{figure*}[ht]
\centering
\begin{tikzpicture}
  \matrix (m) [matrix of math nodes,row sep=3em,column sep=4em,minimum width=2em] {
     \text{QCD at high occupancy} & \text{perturbative QCD} \\
     \text{Gravity at high occupancy} & \text{perturbative gravity} \\};
  \path[-stealth]
    (m-1-1) edge node [left] {\text{Strong field semi-classical double copy}} (m-2-1)
    (m-1-2) edge node [above] {} (m-1-1)
    (m-2-2) edge node [below] {} (m-2-1)
    (m-1-2) edge node [right] {\text{BCJ double copy}} (m-2-2);
\end{tikzpicture}
\caption{Sketch of double copy relations between QCD and gravity in the low and high occupancy regimes of the theories.} \label{Fig1}
\end{figure*}
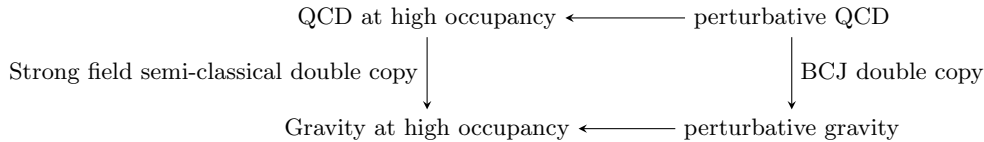

A sketch of the connections outlined is shown in Fig.~\ref{Fig1}. The question we have posed can be sharpened to inquire whether one might take advantage of powerful techniques developed over decades for QCD at high occupancy (following the direction of the arrow at the top of Fig.~\ref{Fig1}) and classical double copy relations (following the path of the arrow at the top left downwards) as a possible route to make progress in the study of gravity at high occupancies relevant for black hole formation at a critical impact parameter $b\sim R_S$. This would be complementary to extending perturbative gravity techniques (following the path of the bottom arrow) directly to the high occupancy regime\footnote{We note that there are significant developments in EFT approaches~\cite{Goldberger:2004jt,Donoghue:2022eay,Beneke:2022pue} to perturbative gravity and in worldline methods~\cite{Jakobsen:2021smu}.}. 

We begin our detailed study with the observation that the QCD Lipatov vertex is simply obtained in shockwave collisions in the CGC EFT in two analytically accessible limits~\cite{Blaizot:2004wu,Gelis:2005pt}: a) the ``dilute-dilute" limit where the saturation scales in two colliding ultrarelativistic nuclei are smaller than the typical transverse momentum of the gluons produced and b) the ``dilute-dense" limit where the saturation scale of one of the colliding nuclei is comparable to the transverse momentum of the emitted gluons but smaller than the saturation scale of the other nucleus. To first provide context to the reader unfamiliar with shockwave collisions in QCD, we should mention that study of gluon shockwave collisions in the context of a first principles approach to Quark-Gluon Plasma (QGP) formation in ultrarelativistic heavy-ion collisions goes back twenty five years. Analytic formulations were first developed in the ``dilute-dilute" regime~\cite{Kovner:1995ja,Kovner:1995ts,Kovchegov:1997ke,Gyulassy:1997vt} and subsequently extended to the ``dilute-dense" regime~\cite{Kovchegov:1998bi,Dumitru:2001ux,Blaizot:2004wu,Gelis:2005pt}. The full ``dense-dense" shockwave collisions most relevant to QGP formulation are only accessible through numerical simulations of Yang-Mills equations initiated in \cite{Krasnitz:1998ns,Krasnitz:2000gz,Lappi:2003bi}. A complementary approach to gluon shockwave collisions at strong coupling in QCD-like theories is that initiated taking advantage of the AdS/CFT correspondence~\cite{Gubser:2008pc,Grumiller:2008va,Chesler:2008hg,Albacete:2009ji}. While fundamentally different from the perspective here, there are nevertheless interesting common technical features that we will comment on in our later discussion. For a summary of subsequent developments and progress in the study of thermalization of the QGP in both weak coupling-strong field and strong coupling frameworks, we refer the reader to \cite{Berges:2020fwq}. 

A key idea in the CGC EFT is that partons at large rapidities with momentum fractions $x\sim 1$ in the $2\rightarrow N$ process are static on the relevant time scales of the scattering and source the production of $x\ll 1$ ``wee" gluons at smaller rapidities. In the semi-classical picture, these large $x$ partons can be represented by a higher dimensional (classical) color representation since dynamical wee gluons couple to a large number of static color charges when $N\gg 1$~\cite{McLerran:1993ka,Jeon:2004rk}. Hence inclusive gluon production at a given rapidity of interest and fixed impact parameter (in $2\rightarrow N$ scattering where the fastest sources have light cone momenta $P^\pm \rightarrow \infty$) is described by solutions of 3+1-D Yang-Mills equations in the presence of two static (independent of light cone $x^\pm$) color charge densities $\rho_{A,B}(\bsx)$ that are $\delta$-functions in $x^{\mp}$. Further, in this infinite momentum frame picture the entire dynamics of the color charge distributions is in 2-D transverse space\footnote{This gives a simple explanation for the 2-D nature of Lipatov's EFT.}. As we will describe at length, it is in the solution of the Yang-Mills equations - in the limits $\rho_A/\square_\perp, \rho_B/\square_\perp \ll 1$ and $\rho_A/\square_\perp \ll 1$ and $\rho_B/\square_\perp \sim 1$ (with $\square_\perp = \delta_{ij}\partial_i \partial_j$) corresponding respectively to the dilute-dilute and dilute-dense frameworks - in which one recovers the Lipatov vertex~\cite{Blaizot:2004wu,Gelis:2005pt}. (Here $\square_\perp$ denotes the typical squared transverse momenta of wee gluons in the scattering of nucleus A off nucleus B.) Albeit very simple, this description in terms of transverse color charge density distributions goes to the heart of the RG~\cite{Jalilian-Marian:1997qno}; the semi-classical picture of $2\rightarrow N$ scattering described as the scattering of shockwaves is reproduced with changing rapidity $Y\rightarrow Y^\prime$ with the dominant quantum fluctuations in the window $\Delta Y = Y^\prime -Y$ absorbed into the evolution of the sources $\rho^{A,B}_Y \rightarrow \rho^{A,B}_{Y^\prime}$. 

In Einstein-Hilbert gravity, we analogously treat $2\rightarrow N$ scattering by modeling the large occupancy of (static on relevant time scales) gravitons above and below the rapidity of interest as possessing densities $\rho_A$, $\rho_B$.  The semi-classical problem is then the scattering of the corresponding gravitational shockwaves and the derivation of the inclusive spectrum of emitted gravitons.  As in the Yang-Mills case, we will employ the dilute-dilute and dilute-dense classification, with the relevant expansion parameters being $R_L/b$ and $R_H/b$, where $R_L$ and $R_H$ are the radii of the regions of overoccupied gravitons and $b$ is the impact parameter in the scattering. In this paper, we will restrict ourselves to the dilute-dilute case $R_L/b, R_H/b \ll 1$ but shall comment on the dilute-dense ($R_L/b\ll 1$, $R_H\sim 1$) scenario which we will address in follow-up work. We will demonstrate explicitly that the Lipatov double copy is recovered in the dilute-dilute framework. 

We note that there is an extensive literature on the classical double copy between Yang-Mills and Einstein-Hilbert gravity, quite some of which is relevant to our work and directly inspired aspects\footnote{The color-kinematic duality between gluon radiation in the dilute-dilute regime computed in \cite{Kovner:1995ja,Kovner:1995ts,Kovchegov:1997ke,Gyulassy:1997vt} to that of gravitational radiation was discussed in \cite{Goldberger:2016iau,Goldberger:2017frp}. Since the presence of the Lipatov vertex in classical Yang-Mills solutions was only first noted in 
\cite{Blaizot:2004wu,Gelis:2005pt} for both dilute-dilute and dilute-dense shockwave collisions, the Lipatov double copy relation was not identified in these works.} of it~\cite{Saotome:2012vy,Akhoury:2013yua,Monteiro:2014cda,Luna:2016due,Goldberger:2016iau,Goldberger:2017frp,Shen:2018ebu,Plefka:2018dpa,Kosower:2018adc}. However as noted, what is meant by classical for the most part in these works is different from what is meant by classical in the deeply inelastic strong field regime of copious graviton radiation where it is the quantum mechanical coherence of radiation that leads to emergent classical behavior for $N\gg 1$ on time scales of interest. 
The outstanding question here is whether black hole formation occurs at some critical impact parameter. While this question has been addressed\footnote{An engaging recent discussion and additional references can be found in \cite{Page:2022bem}.} for  classical gravitation shockwave collisions in several analytical and numerical works~\cite{Eardley:2002re,Giddings:2004xy,Yoshino:2002tx,Yoshino:2005hi,Pretorius:2007jn,Sperhake:2012me} - following pioneering work in ~\cite{DEath:1976bbo,Kovacs:1978eu,DEath:1992plq,DEath:1992mef,DEath:1992nmz,Verlinde:1991iu} - none of these works, to the best of our knowledge, do so in the context of uncovering a semi-classical fixed point at a critical impact parameter of RG evolution. In other words, what is the gravitational equivalent of the BK equation in Regge asymptotics? The work of Lipatov, of ACV et al., and of Dvali et al., are very suggestive,  as is the large body of work we noted on the emergent dynamics of gluon saturation in QCD. We will outline the necessary ingredients here but will leave more detailed investigations to future papers. 

Recently the BCJ double copy has emerged as a powerful tool to study gravitational radiation emerging from the classical conservative dynamics of binary black hole collisions with amplitude computations providing results to $O(G^3)$~\cite{Bern:2019nnu} and recently even $O(G^4)$ in the post-Minkowski expansion~\cite{Bern:2022jvn}\footnote{The references \cite{Damgaard:2023ttc,Damgaard:2023vnx,Bjerrum-Bohr:2021din,Bjerrum-Bohr:2021vuf} include the recent studies on post-Minkowskian expansion in general relativity from modern amplitudes perspective where the subtleties related to the extraction of classical physical observables from quantum mechanical calculations are articulated.}. An excellent discussion of its relation to different approaches in the literature can be found in \cite{Damour:2019lcq}. It is therefore natural to consider potential gravitational wave signatures from the deeply inelastic strong field regime of our interest where radiation effects significantly modify eikonal multiple scattering. A useful connection of our work to the post-Minkowskian analysis is through the the ACV approach that we discussed earlier~\cite{Damour:2017zjx}. We will briefly discuss potential future applications of the Regge EFT we develop here. 

The paper is organized as follows. In Section~\ref{sec:LipatovReview}, we briefly summarize the perturbative computation of the $2\rightarrow N$ ladder in QCD and gravity in multi-Regge asymptotics. 
In Section~\ref{subsec:Lipatov-QCD}, we discuss how one goes from the computation of perturbative amplitudes in the Regge regime  of QCD to a quantitative semi-classical picture, as illustrated by the arrow to the top of Fig.~\ref{Fig1}. In particular, we discuss how gluon radiation can be computed in shockwave collisions in a systematic power counting scheme. In Section~\ref{subsec:Lipatov-gravity}, we outline the extant discussion of perturbative gravity and the Lipatov vertex in the Regge regime and motivate the semi-classical approach suggested by the arrow to the bottom left of Fig.~\ref{Fig1}. 

The gravitational analog of gluon shockwave collisions is discussed at length in Section~\ref{sec:gr-shockwaves}. We begin by introducing the Aichelburg-Sexl shockwave metric for mass distributions with transverse extent and demonstrate the structure of the metric in different coordinate frames. We discuss linearized fluctuations about this shockwave background and the structure of the gravitational Wilson line in light cone gauge. This structure is a gravitational double copy of the identical object in QCD. Shockwave collisions are discussed next albeit only in a dilute-dilute approximation corresponding to impact parameters that are small enough to trigger inelastic graviton production but sufficiently large that multiple scattering/nonlinear (``tidal") effects are sub-dominant. A key element in this derivation is the solution of the geodesic equations for the energy-momentum tensor which admits a contact term  that is essential to 
recover the Lipatov double copy in the solution of the shockwave equations of motion. 
We briefly discuss the relation of our double copy derivation of the Lipatov result to prior classical double copy work in the literature. These double copy connections will be fleshed out further in forthcoming work. 

In Section~\ref{sec:Outlook}, we discuss at some length further directions under investigation that were prompted by the results in this paper. The first of these is the spectrum of gravitational wave radiation and potential observational consequences thereof. The second is the generalization of our dilute-dilute results to the dilute-dense regime where gravitation wave radiation takes place in close vicinity to one of the black holes and therefore necessitates multiple scattering/nonlinear corrections to the spectrum in the ultraviolet. This goes hand-in-hand with the derivation of graviton propagators in the dense shockwave background, whose structure also bears exact analogy to the QCD case. A major future step would be to put all these elements together to derive a renormalization group framework towards understanding black hole formation as a possible fixed point of the evolution at a critical impact parameter. This program is motivated by the QCD case where one understands the emergence of gluon saturation as a nontrivial fixed point of renormalization group evolution in rapidity. We will also discuss in this section a complementary understanding of the classical double copy in the language of asymptotic symmetries and the possible universal features of the Goldstone dynamics of wee partons in QCD and gravity on the celestial sphere. These ideas are motivated by the remarkable exact analogy of a color memory effect in QCD to that in gravity~\cite{Pate:2017vwa}; the former is concretely realized in the CGC EFT~\cite{Ball:2018prg}.

We end the paper with a brief summary. The paper includes several appendices that contain details of the computations in the main text. In Appendix~\ref{appA}, we introduce readers unfamilar with gluon shockwave computations to some of the details of the computations. Most importantly, we show how the QCD Lipatov vertex is derived for dilute-dense shockwave collisions. Appendix~\ref{appB} contains a brief review of general relativity, the conventions employed and many of the details of the linearized Einstein equations presented in the main text. The geodesic solutions for the evolution of the energy-momentum tensor are discussed in Appendix~\ref{appC}. Details of the extraction of the gravitational Lipatov vertex and Fourier transforms are given in Appendix~\ref{appD}. 

\section{\texorpdfstring{$ 2\to N$}{} Gluon and graviton amplitude in multi-Regge kinematics:\\
Lipatov vertex and reggeized propagators}
\label{sec:LipatovReview}
In this section, we will discuss the building blocks of the cascade of wee partons that leads to high occupancy states in QCD and gravity in Regge asymptotics. In addition to the aforementioned 2-D EFT work of Lipatov, the ACV et al. and Dvali et al. approaches, the dynamics of wee gravitons under boosts is a key feature of Susskind's wee parton interpretation~\cite{Susskind:1994vu} of the t'Hooft's holographic principle~\cite{tHooft:1993dmi}. It is therefore useful and important to discuss modern EFT approaches to wee parton dynamics in QCD that in particular address the phenomenon of gluon saturation in the context of further applications of these ideas to the Regge limit of gravity. In Section~\ref{subsec:Lipatov-QCD}, we will provide a brief introduction to these ideas in QCD and shall  motivate the semi-classical framework that formulates 
$2\rightarrow N$ scattering in the language of gluon shockwave collisions. Readers familiar with these ideas can skip this subsection altogether. In Section~\ref{subsec:Lipatov-gravity}, we will outline the parallel formulation of these ideas in gravity (noting both similarities and differences to the QCD case) that will motivate the discussion of gravitational shockwave collisions in Section~\ref{sec:gr-shockwaves}. We will return to some of the ideas\footnote{Note that shockwave solutions for 2-D EFTs of QCD and gravity were also discussed previously in \cite{Verlinde:1991iu,Verlinde:1993te}.} outlined in Section~\ref{subsec:Lipatov-gravity} in Section~\ref{sec:Outlook}.

\subsection{Scattering in QCD in multi-Regge kinematics: from amplitudes to shockwave collisions}
\label{subsec:Lipatov-QCD}

\begin{figure}[ht]
\centering
\includegraphics[scale=1]{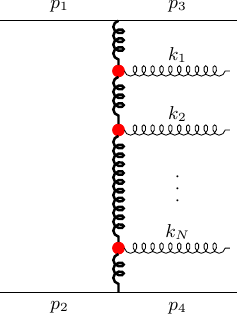}
\caption{Multi-gluon production amplitude in multi-Regge kinematics depicting the two key components in BFKL evolution: (1) The red blobs represent the nonlocal Lipatov's effective vertex, and (2) the thick vertical gluon lines represent the reggeized gluon propagator incorporating (leading logarithmic in $x$) all-order virtual corrections.}
\label{fig:2}
\end{figure}

The Lipatov vertex in QCD is a fundamental building block of the BFKL equation describing the high-energy evolution of QCD cross-sections.  The other essential component is the reggeized gluon propagator. Together, as illustrated in Fig.~\ref{fig:2}, they respectively constitute the real and virtual parts of the  kernel $\mathcal{K_{\rm BFKL}}$ of the BFKL Hamiltonian describing the $2\rightarrow N$ scattering amplitude in Regge asymptotics. We will sketch the key elements here and will refer the reader to the reviews \cite{DelDuca:1995hf,DelDuca:2022skz}
for more details. 

Consider  $2\to 2$ scattering of hadrons in the Regge limit where $s=(p_1+p_2)^2\to \infty$ and $t=(p_1-p_3)^2$ held fixed, with $|t| \gg \Lambda_{\rm QCD}^2$. This scattering regime is characterized by strong ordering in the light cone momenta of the two final state particles: $p_3^+\gg p_4^+$ with $|\bsp_3| \simeq |\bsp_4|$ which is equivalent to strong ordering in their rapidities $y$\footnote{Light cone momenta can be parametrized by their rapidity $y$ as $p=(p^{+} = |\bsp| e^{+ y},p^{-} = |\bsp| e^{- y},\bsp)$.}. When there are $N$ additional soft gluon emissions, as shown in Fig.~\ref{fig:2}, the corresponding multi-Regge kinematics (MRK) is specified by strong ordering of these gluons in their respective rapidities, with the  transverse momenta $\bsk$ of all final state particles being of comparable magnitude:
\be
\label{MRKdef}
y_0^+ \gg y_1^+ \gg y_2^+ \gg \cdots \gg y_N^+ \gg y_{N+1}^+~ \qquad \text{with}~ \qquad \bsk_i \simeq \bsk~.
\ee
We labeled here the rapidities of particle (3) and (4) in Fig. \ref{fig:2} as $y_0$ and $y_{N+1}$ respectively. At energies much higher than any mass scale in the theory, the dominant contributions to inelastic $2\to 2+N$ processes comes from the kinematic region specified in Eq.~\eqref{MRKdef}.

The aforementioned BFKL equation predicts a rapid growth of deeply inelastic scattering (DIS)  electron-hadron scattering  cross-section at high energies such as those accessed at the HERA collider and the future Electron-Ion Collider (EIC)~\cite{Aschenauer:2017jsk}. In DIS, the hard scale is the spacelike square momentum $Q^2$ of the exchanged virtual photon. The BFKL equation can be used to compute the inclusive DIS cross-section at small $x_{\rm Bj}$ (where $x_{\rm Bj} \sim Q^2/s$)  whose energy evolution is characterized by the unintegrated gluon distribution $\mathcal{F}(x,\bsk)$, with $|\bsk|\sim \sqrt{Q^2}$, the typical transverse momenta of emitted gluons. At small values of gluon momentum fractions $x\sim x_{\rm Bj}$, $\mathcal{F}(x,k)$ satisfies the evolution equation 
\be
\frac{\partial \mathcal{F}}{\partial \log (1/x)} = \mathcal{K_{\rm BFKL}} \otimes \mathcal{F}~,
\label{BFKL}
\ee
where $\mathcal{K_{\rm BFKL}}$ is the BFKL kernel and $\otimes$ represents a convolution in transverse momenta. At leading logarithmic accuracy in $x$ (LLx) , $\alpha_s \log(1/x))\sim 1$, with $\alpha_s$  the strong coupling constant, this RG evolution equation resums all-order $(\alpha_s \ln(1/x))^n$ contributions in perturbation theory. This in turn gives rise to a rapid monotonic rise of the inclusive DIS cross-section as $\sigma \sim s^{\omega}~,$ where the exponent $\omega = 4N_c\alpha_s(\log 2)/\pi$ and $N_c$ is the number of colors\footnote{This growth in the cross-section is far more rapid than observed at HERA. 
It is tamed both by including next-to-leading logarithmic contributions and by gluon saturation effects~\cite{Morreale:2021pnn} as we will discuss further.}.

 The precise details of the kernel $\mathcal{K}_{\rm BFKL}$ can be found  for instance in \cite{Kovchegov:2012mbw}; they are not relevant at this stage of our discussion to the corresponding dynamics in gravity. However its building blocks are important to understand and we will briefly sketch the structure of the virtual and real contributions to the kernel. 

\begin{figure}[ht]
\centering
\raisebox{-27pt}{\includegraphics[scale=1]{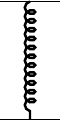} }
=
\raisebox{-27pt}{\includegraphics[scale=1]{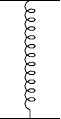} }
+
\raisebox{-27pt}{\includegraphics[scale=1]{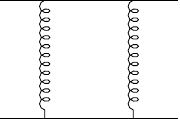} }
+
\raisebox{-27pt}{\includegraphics[scale=1]{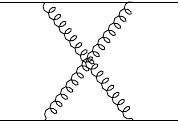} }
+ $\cdots$
\caption{The reggeized gluon, illustrated as a thicker version of the lighter bare gluon, resums all of the leading $\log (s/|t|) \log (|t|/\lambda)$ corrections to the single gluon exchange amplitude, where $\lambda$ is an infrared cutoff that cancels in cross-sections.} 
\label{fig:3}
\end{figure}

The virtual part of the BFKL kernel $\mathcal{K}_{\rm BFKL}$ comes from multiple t-channel gluon exchange diagrams which exponentiate in the regime given by  Eq.~\eqref{MRKdef}  \cite{Lipatov:1976zz}. The reason for exponentiation is that in the MRK regime every additional virtual gluon exchange, as shown Figure \ref{fig:3}, is accompanied by a leading logarithmic factor of $\alpha_s \log(s/|t|)\,\alpha(t)$. Here $\alpha(t)$ is  the one-loop gluon Regge trajectory defined to be 
\be
\label{reggeizedGluon}
\alpha(t) = \alpha_s N_c\,t \int \frac{d\bsk}{(2\pi)^2}\frac{1}{\bsk^2 (\bsq-\bsk)^2}~,~~~~t=-\bsq^2~.
\ee
(Note that cross-diagrams such as 
those shown in  Fig. \ref{fig:2} are suppressed in this kinematics.) 
Since $s/|t|\rightarrow \infty$ in Regge kinematics, $(\alpha_s \log(s/|t|))^n \sim O(1)$ to all n-orders, which allows for the exponentiation of all such contributions. This gives rise to the reggeized gluon propagator where one simply replaces the $i$'th gluon propagator in Fig. \ref{fig:2} by
\be
\frac{1}{t_i} \to \frac{1}{t_i} e^{\alpha(t_i)(y_{i-1}-y_i)}\,.
\ee
The difference $y_{i-1}-y_i$ in the rapidities of the $i-1$ and $i$'th final state particles is proportional to the Mandelstam variable $\log s_{i-1,i}$. Consequently, the replacement in Eq.~\eqref{reggeizedGluon} is proportional to $s^{\alpha(t)}$ . This can be viewed as describing the exchange of a ``quasi-particle" of spin $j=1+\alpha(t)$ -- the reggeized gluon -- depicted in Fig. \ref{fig:3} as a thick gluon line. Reggeization in QCD has been shown to hold to next-to-leading-log accuracy but breaks down beyond. For the state-of-the art, we refer the reader to \cite{DelDuca:2022skz}. 

The real part of $\mathcal{K}_{\rm BFKL}$ comes from the square of Lipatov's central emission effective vertex $C_\mu$ which is depicted as a red dot in Fig.~\ref{fig:2}. This vertex describes the production amplitude of a soft gluon in the scattering of partons in Regge kinematics. Fundamentally, one can use the Jacobi identity in MRK kinematics to represent the sum of the four Feynman diagrams that correspond to bremsstrahlung emission from initial and final state hadrons plus the diagram with an emission from internal virtual process - as a diagram with one effective emission vertex - as shown in  Fig.~\ref{fig:4}:

\begin{figure}[ht]
\centering
\raisebox{-27pt}{\includegraphics[scale=1]{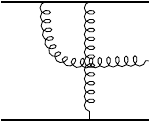} }
+
\raisebox{-27pt}{\includegraphics[scale=1]{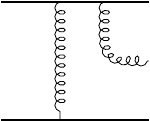} }
+
\raisebox{-27pt}{\includegraphics[scale=1]{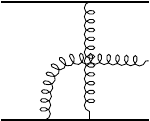} }
+
\raisebox{-27pt}{\includegraphics[scale=1]{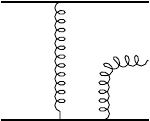} }
+
\raisebox{-27pt}{\includegraphics[scale=1]{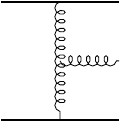} }
\vspace{.25in}

=
\raisebox{-58pt}{\includegraphics[scale=1]{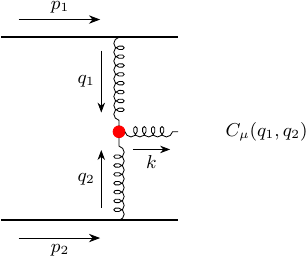} }
\caption{Soft gluon emission diagrams contributing to the Lipatov effective vertex $C_\m(\bsq_1,\bsq_2)$ denoted by the red dot. (Contributions from u-channel crossed diagrams are suppressed as $|t|/s$.)} 
\label{fig:4}
\end{figure}

Adding up the five diagrams in the MRK regime, one finds the following expression for the effective vertex $C_\m$: 
\begin{align}
\label{QLV}
C_\m(\bsq_1,\bsq_2)\simeq -\bsq_{1\m}+\bsq_{2\m} + p_{1\m}\left(\frac{p_2 \cdot k}{p_1 \cdot p_2}-\frac{\bsq_1^2}{p_1 \cdot k}\right)-p_{2\m}\left(\frac{p_1 \cdot k}{p_1 \cdot p_2}-\frac{\bsq_2^2}{p_2 \cdot k}\right)~,
\end{align}
where the $\simeq$ sign indicates that only the transverse components of the momenta $q_1,q_2$ of exchanged gluons are relevant in  MRK kinematics. The vertex $C_\mu$ is a function of the transverse momenta $\bsq_1, \bsq_2$ alone; despite the form of the equation above, it has no dependence upon the incoming external momenta $p_1, p_2$. (This will become apparent in light cone gauge as we will describe shortly.)  Indeed, the Lipatov vertex is universal in the sense that it is insensitive to the nature of the external particles (such as for instance its spin). Also, it is gauge covariant; one can check that the Ward identity $k^\mu C_\m =0$ holds, where $k=q_1+q_2$. This result for the effective emission vertex was first obtained in QCD by Lipatov-hence the name \cite{Lipatov:1976zz}. As we noted, and shall discuss shortly, its analog in gravity for $2\rightarrow 3$ was also computed by Lipatov\footnote{It was further generalized in the context of open string scattering in \cite{Ademollo:1989ag,Ademollo:1990sd} which in the limit $\alpha'\to 0$ agrees with Eq.~\eqref{QLV}.}.

The outstanding achievement of Lipatov and colleagues was to show that $2\rightarrow N$
scattering (or equivalently, the imaginary part of the $2\rightarrow 2$ amplitude) in MRK kinematics could in leading logarithmic kinematics be described as an iteration of one rung of the ladder containing reggeized gluons and the Lipatov vertex, as illustrated in Fig.~\ref{fig:2}. The color singlet projection of the exchange of two reggeized gluons is the perturbative Pomeron, the weak coupling counterpart of the soft Pomeron often invoked to describe the systematics of total cross-sections~\cite{Donnachie:2013xia}. 

However BFKL dynamics is not the complete story. Firstly, the solutions of the BFKL equation show the unintegrated gluon distribution $\mathcal{F}(x,\bsk)$ diffuses to  infrared and ultraviolet momenta with increasing rapidity. The former is clearly troublesome since that's the nonperturbative regime ${\bf k}\sim \Lambda_{\rm QCD}$ where weak coupling computations are invalid. Further, the rapid growth of the inclusive cross-section for a fixed impact parameter violates unitarity at large rapidites. Not least, the increasing phase space occupancy due to the rapid proliferation of gluons at small $x$ suggests that many-body (higher twist) correlations are important. All of these issues persist at next-to-leading logarithmic accuracy. 

Perturbative and nonperturbative arguments suggest that the phase space occupancy of gluons in QCD can maximally be $O(1/\alpha_s)$, leading to a much weaker growth of gluon distributions. This phenomenon, known as gluon saturation~\cite{Gribov:1983ivg,Mueller:1985wy}, is characterized by an emergent close packing scale $Q_S(x)\gg \Lambda_{\rm QCD}$ at maximal occupancy that is responsible for the unitarization of the inclusive cross-section at fixed impact pararameter. In other words, for a weakly interacting probe of given fixed $Q^2$ with $\alpha(Q^2)\ll 1$, there is a corresponding value of $x$ for which the probe scatters of the hadron target with unit probability at occupancy $N\sim 1/\alpha_s\gg 1$.

\begin{figure}[ht]
\centering
\includegraphics[scale=1]{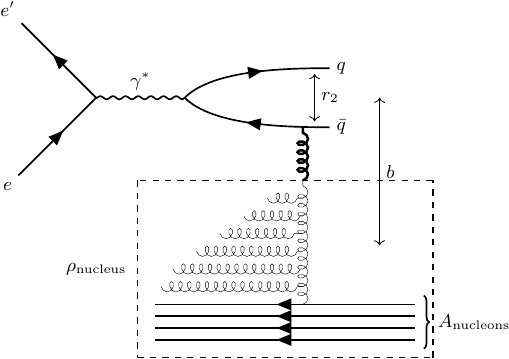} 
\caption{The scattering of a quark-antiquark dipole in DIS off a boosted heavy nucleus. The length of the emitted gluon lines indicate their distance in rapidity from ``valence" partons whose rapidities are close to the light cone of the heavy nucleus. At high energies, partons inside the dashed box can be represented by a coherent color charge density $\rho_{\rm nucleus}$ that is static on the times scales of the interaction. }
\label{dipole-interaction}
\end{figure}

We noted earlier that small $x$ physics at high occupancy is described by the Color Glass Condensate (CGC) EFT~\cite{Gelis:2010nm,Iancu:2003xm,Kovchegov:2012mbw}. The underlying physics in the DIS context is captured by the illustration of the DIS process in 
Fig.~\ref{dipole-interaction}. In the rest frame of the dipole target, the fast moving nucleus with momentum $P^+\rightarrow \infty$ emits a large number of gluons which in the Regge limit are ordered in rapidity, with the fastest gluons (represented by the longest gluon lines) co-moving with the valence degrees of freedom on the light cone and the slowest (small $x$) gluons scattering off the target, with $x\sim x_{\rm Bj}$. 

A dipole of size $r_2 \sim 1/Q$, at an impact parameter 
$b \ll 1/\Lambda_{\rm QCD}$ will exchange colored gluons with a lump (represented by the 
dotted rectangle) of maximal size $1/Q_S$ consisting of static color sources on the relevant time scale of the scattering. Since the lump has a high occupancy $N\sim 1/\alpha_s$, it is quasi-classical, with energy levels\footnote{Since the energy separations of the high occupancy screened gluons are $\sim Q_S/N$, the characteristic decay time of the shock is $\sim \frac{1}{\alpha_s Q_s}$  which is considerably longer than the typical resolution scale $1/Q$ of the probe but much shorter than the eikonal time scale $\sim P^+/Q^2$~\cite{Dvali:2021ooc}. } separated by $1/N$; further, since the lump contains a large number of color charges, its color charge is given by a higher dimensional classical representation of the $SU(N_c=3)$ algebra~\cite{McLerran:1993ka,McLerran:1993ni,Jeon:2004rk}. 

Since the dipole perceives the lump as being Lorentz contracted in its rest frame, the lumpy ``shockwave" distribution can be represented by a transverse color charge density distribution $\rho_{\rm N}(\bsx)$. The dipole scattering at leading order in this EFT picture is then captured by the formula 
\be
\langle d\sigma_{\rm LO}\rangle = \int [D\rho_{\rm N}]\, W_{Y_0}[\rho_{\rm N}] 
\,d{\tilde \sigma}_{\rm LO}[\rho_{\rm N}]\,.
\label{eq:CGC}
\ee
In this expression, $d{\tilde \sigma}_{\rm LO}$ denotes the tree level scattering amplitude of the quark-antiquark dipole off the classical ``Coulomb" background field generated by the source distribution $\rho_{\rm N}(\bsx)$ and $W_{\rm Y_0}[\rho_{\rm N}]$ is the gauge invariant stochastic distribution of these color sources at the rapidity scale $Y_0$ separating the sources from the fields that interact with the probe. At next-to-leading order, including the leading $O(\alpha_s)\Delta Y$ real and virtual corrections to the shockwave with the change of rapidity $\Delta Y= Y_1-Y_0$, one obtains the same structure as Eq.~\eqref{eq:CGC} for the evolved lump distribution $\rho_{\rm N}[Y_0]\rightarrow \rho_{\rm N}[Y_1]$, with $W_{\rm Y_1}[\rho_{\rm N}]$ satisfying the Wilsonian renormalization group equation
\be
\frac{\partial W_{\rm Y_1}[\rho_{\rm N}]}{\partial \Delta Y} = {\cal H}_{\rm JIMWLK}\, W_{\rm Y_1}[\rho_{\rm N}]\, .
\ee 
An essential ingredient in this computation is the computation of quark and gluon shockwave propagators~\cite{McLerran:1994vd,Ayala:1995kg,Ayala:1995hx,Balitsky:1995ub,McLerran:1998nk,Balitsky:2001mr} that can be mapped to gluon-gluon-reggeized gluon (and quark-quark-reggeized gluon) propagators~\cite{Bondarenko:2017ern,Hentschinski:2018rrf} in Lipatov's reggeon field theory that we alluded to previousy. These have their counterpart in the effective vertex for the graviton-reggeized graviton-reggeized graviton three-point vertex that we will return to later in the paper. 

The JIMWLK Hamiltonian~\cite{Jalilian-Marian:1997qno,Jalilian-Marian:1997ubg,Jalilian-Marian:1998tzv,Iancu:2000hn,Iancu:2001ad,Ferreiro:2001qy} has an extremely rich structure corresponding to an n-body hierarchy of lightlike Wilson line correlators~\cite{Balitsky:1995ub} that describe arbitrary final states in the CGC EFT to LLx accuracy\footnote{This RG framework extends to NLLx accuracy; the discussion of these developments is however outside the scope of this manuscript.}. In particular, the 
inclusive dipole correlator $d\sigma_{\rm LO}\propto 2 N_c [1-S]$ where the dipole S-matrix $S=\frac{1}{N_c}\langle V(x_\perp) V^\dagger(y_\perp)\rangle_{\rho_{\rm N}}$. Here  $V(x_\perp) = P\exp(i\frac{\rho_{\rm N}}{\square_\perp^2})$ and $x_\perp,y_\perp$ are the dipole coordinates with $r_2 = x_\perp - y_\perp$ and $b = (x_\perp+y_\perp)/2$ (for configurations where the quark and antiquark equally share the virtual photon's light cone momentum).  In the large $N_c$ limit, and for a large nucleus with atomic number $A \gg 1$, the JIMWLK equation for the  dipole S-matrix results in the nonlinear BK equation, 
\be
\frac{\partial S}{\partial \Delta Y} = \frac{\alpha_s N_c}{2\pi^2}\mathcal{K_{\rm BFKL}}\otimes [S\,S -S] \,.
\label{eq:BFKL1}
\ee
We have suppressed here the nontrivial coordinate dependence of the quantities on the r.h.s of this equation. In the perturbative limit of $S = 1-{\cal \tilde{F}}$ with ${\cal \tilde {F}}\ll 1$ the Bessel-Fourier transform of the unintegrated gluon distribution, this equation can be linearized and is precisely the coordinate space counterpart of the BFKL equation in Eq.~\eqref{BFKL}. However as is transparent, this equation has in addition a nontrivial fixed point at $S\rightarrow 0$, corresponding to the classicalization and unitarization of the cross-section. 

The relevant message from the above discussion is that the semi-classical framework of 
static sources and dynamical fields in the CGC EFT reproduces the $2\rightarrow N$ scattering amplitude described by the BFKL equation in multi-Regge kinematics, and its nonlinear generalizations. In Regge language, the BK equation resums so-called ``fan" diagrams containing multi-Pomeron interactions. A significant advantage of this approach is that it provides a powerful way to describe multi-particle final states in hadron/nucleus scattering in Regge asymptotics. We will turn now to discuss to a discussion of these collisions and demonstrate how the Lipatov vertex emerges in this framework. 

In the semi-classical framework of the CGC EFT, multi-particle production in Regge asymptotics is described to lowest order by the collision of gluon shockwaves. A natural formalism in this strong field context is the ``in-in" Schwinger-Keldysh formalism as opposed to the ``in-out" formalism of the S-matrix. One considers instead single-inclusive  and multi-particle correlations of the produced gluons rather than the $2\rightarrow N$ scattering probability\footnote{For a detailed discussion of the relation between the ``in-in" and ``in-out" formalism in the context of multi-particle production in quantum field theory, we refer the reader to \cite{Gelis:2006yv,Gelis:2006cr}. Many features of 
the reggeon field theory language of Pomerons and Reggeons can be understood in terms of the combinatorics of cut and uncut sub-graphs (Cutkosky rules) of multi-particle final states in the presence of strong fields~\cite{Gelis:2006ye}. In strong fields, multi-particle probabilities and multi-particle inclusive distributions are qualitatively different objects. See also \cite{Guiot:2020pku} for a recent related discussion.}. For simplicity, we will consider only the single-inclusive gluon (and graviton) ``bremsstrahlung" spectrum. 

For strong sources comprising the large $x$ modes of the scattering nuclei 
(which are order $\rho_{\rm nucleus}\sim 1/g$), the leading term in the power counting is the produced classical field $A_{\rm cl}^\mu$, which too is of order $O(1/g)$; the single-inclusive distribution is therefore of $O(1/\alpha_s)$. At next-to-leading order, there are two sorts of contributions: these are a) the one loop correction to the classical field $a_{\rm quant}^\mu$ which is $O(1)$, and 
b) the small fluctuation propagator $\langle a_{\rm quant}^\mu a_{\rm quant}^\nu\rangle$. 
It is the logarithmic enhancements $\alpha_s \ln(1/x) \sim O(1)$ to these contributions that contribute to the JIMWLK Hamiltonian and are thereby absorbed in the evolution of the single-inclusive gluon distribution. Thus at each step in the rapidity evolution of the individual nuclei, the problem of n-particle inclusive gluon production at a given rapidity is simply the solution of the QCD Yang-Mills equations in the presence of the static source distributions of each of the nuclei evolved up to that scale\footnote{This presumes that the wee partons of each of the nuclei don't talk to each other before the collision - the 
weight functionals $W[\rho_{\rm nucleus}]$ containing the nonperturbative information on their n-body distributions factorize in the collision. This factorization holds when 
$\rho_{\rm nucleus}\sim 1/g$ to LLx accuracy~\cite{Gelis:2008rw,Gelis:2008ad}.}. 

The Yang-Mills (YM) equations for the general problem of shockwave collisions 
is given by 
\begin{align}
\label{YM:dense-dense}
D_\m F^{\m\n}=J_{HI}^\n~,
\end{align}
where $F_{\m\n}=\p_\m A_\n-\p_\n A_\m+ig[A_\m,A_\n]$ is the field strength tensor and $J_{\rm HI}^\m$ is the covariantly conserved current: $ D_\m J_{HI}^\m = 0$. For the case of shockwave scattering of nuclei (as in heavy-ion collisions at ultrarelativistic energies), the shockwave currents can be represented as 
\be
J_{HI}^{\n,a} = \delta^{\n+}\rho^a_A(x_\perp)\delta(x^-) + \delta^{\n-}\rho^a_B(x_\perp) \delta(x^+)
\,.
\ee
Here $\rho^a_A(x_\perp)$ and $\rho^a_B(x_\perp)$ are the quasi-classical color charge distributions of each of the nuclei corresponding to a higher dimensional representation of the color charges depicted in Fig.~\ref{dipole-interaction}. (We emphasize that their RG evolution includes both real emissions and virtual loops to all orders to LLx accuracy.)
These are distributed in the transverse plane of the scattering; for $A\gg 1$, the weight functional\footnote{Unless required, we will henceforth drop the rapidity label $Y$; it should be understood that the classical equations are describing scattering generating a gluon distribution at a given rapidity. } $W[\rho_{A,B}]$ is Gaussian distributed such that $\langle \rho_{A,B}^a(x_\perp)\rho_{A,B}^b(y_\perp)\rangle = Q_S^2\, \delta^{ab}\, \delta^{(2)}(x_\perp-y_\perp)$, with $Q_S^2 \propto A^{1/3}\, \Lambda_{\rm QCD}^2$. Note that the $A\gg 1$ limit of QCD provides an explicit construction~\cite{McLerran:1993ka,McLerran:1993ni,Jeon:2004rk} demonstrating the emergent saturation scale\footnote{For simplicity, we have assumed here that the nuclei are identical. For nuclei with different atomic numbers, there will be two saturation scales reflecting the fact that one has different initial distributions of color charge in each.} we discussed in the introduction. The $\delta(x^\mp)$ terms represent eikonal currents for which classical sub-eikonal corrections are $O(1/P^\pm)$ respectively. 
Finally, observe that the currents are independent of the light cone times $x^\pm$, respectively; this reflects that they are static color sources on the time scales of gluon production at the rapidity of interest. 

The nucleus-nucleus scattering problem thus formulated~\cite{Kovner:1995ja,Kovner:1995ts}  in full generality for $\rho_A\sim 1/g$ can only be solved numerically~\cite{Krasnitz:1998ns,Krasnitz:2000gz,Berges:2020fwq}. However one can identify the expansion parameters $\rho_A/\square_\perp, \rho_B/\square_\perp$ ($\square_\perp\equiv \partial_\perp^2$) in the YM equations that one can expand in to obtain analytic solutions. These are the dilute-dilute YM asympotics of $\rho_A/\square_\perp, \rho_B/\square_\perp \ll 1$ (corresponding to the regime of large transverse momenta $k_\perp \gg Q_S$)~\cite{Kovner:1995ja,Kovner:1995ts,Kovchegov:1997ke,Gyulassy:1997vt} and dilute-dense asymptotics  $\rho_A/\square_\perp\ll 1, \rho_B/\square_\perp \sim 1$~\cite{Dumitru:2001ux,Blaizot:2004wu,Gelis:2005pt}, or $Q_{S,A}\ll k_\perp \ll Q_{S,B}$. 
The dense-dense regime of $\rho_A/\square_\perp, \rho_B/\square_\perp \sim 1$, as noted previously, is not analytically tractable and corresponds to fully nonlinear solutions of the YM equations. 

\begin{figure}[ht]
\centering
\includegraphics[scale=1]{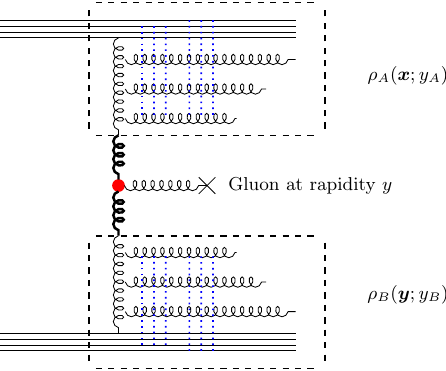} 
\caption{The dilute-dilute regime of shockwave scattering in QCD. The inclusive gluon 
distribution (depicted by the emission of a gluon line at rapidity y)  is insensitive to the eikonal exchanges (depicted with blue-dashed lines) within the color $\rho_A$ and $\rho_B$ that interact via the emission of reggeized gluons that interact via the effective Lipatov vertex.}
\label{dilute-dilute}
\end{figure}

\begin{figure}[ht]
\centering
\includegraphics[scale=1]{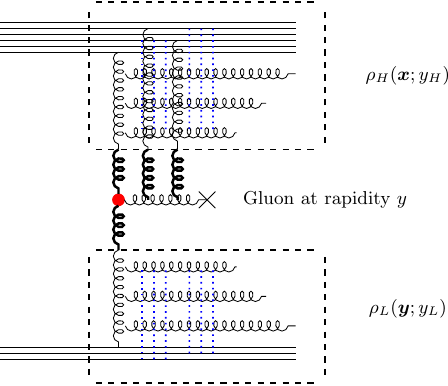}
\caption{Dilute-dense scattering where $\rho_L/\square_\perp \ll 1$ and $\rho_H/\square_\perp\sim 1$. For the latter, coherent multiple scatterings from the nucleus on to the emitted gluon are unsuppressed, and can be summed up into a lightlike Wilson line. }
\label{dilute-dense}
\end{figure}

The dilute-dense scattering case is illustrated in Fig.~\ref{dilute-dilute} and the dilute-dense case in Fig.~\ref{dilute-dense}. In the former case, since $\rho_A/\square_\perp, \rho_B/\square_\perp \ll 1$, coherent multiple scattering is suppressed in both of the colliding nuclei. One can consider this the ``BFKL regime" of high energy scattering since in this limit the energy evolution of  both ``lumps" is described by the BFKL equation. In contrast, in the dilute-dense case (to keep track of light and heavy sources, we shall switch notations here with $\rho_{L,H}$ respectively denoting the corresponding distributions) when $\rho_H/\square_\perp\sim 1$ , the multiple scattering insertions on to the emitted gluon can be absorbed into a Wilson line, as we shall now discuss\footnote{It is not feasible to factorize coherent multiple scatterings from both nuclei into separate Wilson lines. The reader should also note that in this classification one {\it always} has $\rho_{A,B}\sim 1/g$. The situation when $\rho_{A,B} \ll 1/g$ is quite subtle~\cite{Iancu:2004iy,Kovner:2005uw} and beyond the scope of this discussion.}.

In the context of dilute-dilute and dilute-dense scattering, the classical YM problem can be cast as follows. A gluon shockwave with transverse source distribution $\rho_H(\bsx)$ moving in the positive $z$ direction is generated by the current
\begin{equation}
J_{\mu}=g \delta_{\mu-} \delta\left(x^{-}\right) \rho_H(\bsx) ~,
\end{equation}
where $T^a$ is a generator of the color algebra and $\delta(x^-)$ is the Dirac delta function. It is straightforward to verify that the exact solution to YM equations with this current\footnote{Covariant current conservation follows from the equation of motion and the Jacobi identity of the Lie algebra generators. The covariant derivative action on an adjoint field $F$ is given by $D_\m F = \p_\m F - ig [A_\m, F]$.} is given by
\be
\label{gShockBgnd}
\bar{A}_\m(x^-,\bsx) = -g \delta_{\mu-} \delta\left(x^{-}\right) \frac{\rho_{H}\left(\boldsymbol{x}\right)}{\square_\perp}~.
\ee
This is the singular shockwave solution where there is a delta function singularity in $x^-$. It shows that in the regions $x^->0$ or $x^-<0$ the field strength $F_{\m\n}$ vanishes. However as shown below, the gauge fields do not identically vanish. In the region $x^-<0$ the gauge field is trivial but for $x^->0$, it is a pure gauge. This is best seen by performing a gauge transformation to light cone gauge: $A_\m \to U A_\m U^\dagger +\frac{i}{g} U\p_\m U^\dagger$ on the solution in Eq.~\eqref{gShockBgnd} with the transformation matrix $U$ given by
\be
U(x^-,\bsx) = \exp\(ig^2 \Theta(x^-)\frac{\rho_{H}(\bsx)}{\square_\perp}\)~,
\ee
where $\Theta(x^-)$ is the step function. In the new gauge, only transverse components of the gauge field are nonvanishing and are given by
\be
A_i = \frac{i}{g} \Theta(x^-) \tilde{U}\p_i \tilde{U}^\dagger \qquad \text{where} \qquad \tilde{U} = \exp\(ig^2 \frac{\rho_{H}(\bsx)}{\square_\perp}\)~.
\ee
The form of the gluon shockwave clearly demonstrates that in the region $x^->0$, the gauge field is a pure gauge. This shockwave solution is the non-Abelian equivalent of the Weizs\"{a}cker-Williams distribution in classical electrodynamics~\cite{McLerran:1993ka,McLerran:1993ni}.

To set up the shockwave collision problem, one turns on the current of the incoming shockwave with the transverse color charge distribution $\rho_L(\bsx)$ moving in the negative $z$ direction:
\be
J_{\mu} = g\delta_{\mu+} \delta(x^+)\rho_L\(\bsx\)~.
\ee

\begin{figure}[ht]
\centering
\includegraphics[scale=1]{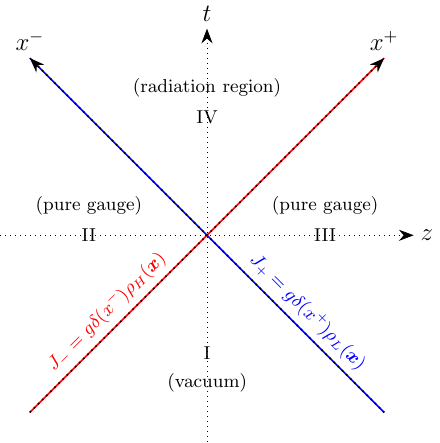}
\caption{Spacetime diagram of collision of two gluon shockwaves. The red and blue lines represent the lightlike trajectories of the two incoming shockwaves. The future light cone of the collision point $t=z=0$ is where radiation occurs. In the other regions, where the field strength vanishes, the gauge field is a pure gauge in regions II and III and trivial in region I.}
\label{spacetimeg}
\end{figure}

Past the collision time $t=0$, in the dilute-dilute approximation, one simply linearizes the YM equations to linear order in the sources $\rho_H$ and $\rho_L$ and solves\footnote{In the QCD case, in marked contrast to gravity, the impact parameter (in transverse space) between the two shockwaves has to be much smaller than the size of the hadron to ensure the gauge coupling is small enough to apply perturbative methods to compute the radiation field $a_\mu$ from classical YM equations.} for the radiation field $a_\mu$. In light cone gauge $a_+=0$, the result 
(derived in detail in Appendix \ref{appA}) for the physical components of the gauge field takes the form 
\begin{align}
    &\square a_{i,c} = -g^3\(\Theta(x^+)\Theta(x^-) \p_i\(\frac{\rho_H}{\square_\perp}\rho_L\)-2\delta(x^+)\delta(x^-) \frac{\rho_H}{\square_\perp} \frac{\p_i\rho_L}{\square_\perp}\)T^a T^b f_{abc}~.
\end{align}
Taking the Fourier transform of this equation and putting the momenta of the emitted gluons  on-shell $k^2 = 2k_+k_--\bsk^2=0$, one obtains 
\begin{align}
    \label{gaugeLV}
    a_{i,c}(k) &= -\frac{2ig^3}{k^2+i\epsilon k^-}\int \frac{d^2\bsq_2}{(2\pi)^2}\(q_{2i}-k_i\frac{\bsq_2^2}{\bsk^2}\)\frac{\rho_H}{\bsq_1^2}\frac{\rho_L}{\bsq_2^2}T^a T^b f_{abc}~.
\end{align}
In this result for the radiation field,  $1/k^2$ corresponds to the emitted gluon propagator,  $1/\bsq_1^2$ and $1/\bsq_2^2$ are the exchanged reggeized gluon propagators and the Lipatov vertex appears as the term in the parenthesis, as illustrated in Fig.~\ref{dilute-dilute}. 

To see the latter, we will recast the covariant expression for the Lipatov vertex in Eq.~\eqref{QLV} into light cone gauge. Towards this end, we partially gauge fix the gluon polarization vector $\varepsilon_+=0$. This implies $\ep_-k^-=\ep_ik_i~,\ep_\m p_1^\m =0~,\ep_\m p_2^\m=\ep_-p_2^-$ since $p_1 = (p_1^+,0,0,0)$ and $p_2 = (0,p_2^-,0,0)$. With these relations, it is straightforward to deduce the form of $C^\mu$ in light cone gauge \cite{Ioffe:2010zz}
\begin{align}
\label{QCDLVLC1}
&\ep^*_{\m}(k) C^\m(q_1,q_2) = -2\ep^*_i \(q_{2i}-k_i\frac{\bsq_{2 }^{2}}{\bsk^{2}}\) \equiv \ep^*_i C_i(\bsq_1,\bsq_2)~.
\end{align}
The light cone gauge expression makes transparent the fact that the dependence of this vertex is only on $\bsq_1$ and $\bsq_2$ and not the external momenta $p_1,p_2$.

In \cite{Blaizot:2004wu}, Blaizot, Gelis and one of us, and later Gelis and Mehtar-Tani in \cite{Gelis:2005pt}, noted that the Lipatov vertex is embedded the classical YM solutions in both dilute-dilute and dilute-dense scattering regimes. The result of the dilute-dense computation gives 
\begin{align}
\label{dilutedenseQCD}
a_i(k)  =  -\frac{2ig}{k^2+i\epsilon k^-}\int \frac{d^2\bsq_{2}}{(2\pi)^2} \(q_{2i}-k_i\frac{\bsq_2^2}{\bsk^2}\) \frac{\rho_L(\bsq_{2})}{\bsq_{2}^2}\bigg(U(\bsk+\bsq_{2})-(2\pi)^2 \delta^2(\bsk+\bsq_{2})\bigg)\,,
\end{align}
where $U(\bsk)$  is the Fourier transform of the lightlike Wilson line operator
\begin{align}
    U(x^-, \bsx)  \delta\left(x^{+}\right)  = \exp\(ig \int_{-\infty}^{x^-} dz^- \bar{A}_-(z^-, \bsx) \cdot T \)~,
\end{align}
where $\bar{A}_-(z^-, \bsx) $ was given in Eq.~\eqref{gShockBgnd}. As observed earlier, the Wilson line encodes the coherent multiple scattering of the emitted gluon off the dense source $\rho_H$ in Fig. \ref{dilute-dense}. Expanding the above result to lowest order in $\rho_H$ allows one to recover the dilute-dilute result in Eq.~\eqref{gaugeLV}.

To summarize, the Lipatov vertex first computed in the context of the $2\rightarrow 3$ scattering amplitude can be obtained from solutions of the classical YM equations in the presence of nontrivial sources that evolve with rapidity via the BFKL/BK/JIMWLK equations depending on the kinematics of interest. As emphasized,  an important ingredient in this derivation are the shockwave propagators in the strong background fields $\rho_{A,B}\sim 1/g$. Reggeized gluons can be understood as the gauge fields coupling to these sources~\cite{Jalilian-Marian:2000pwi,Caron-Huot:2013fea}; we will return to this discussion in Section~\ref{sec:Outlook}.

\subsection{Scattering in gravity in multi-Regge kinematics: from amplitudes to shockwave collisions.}
\label{subsec:Lipatov-gravity}

In Section~\ref{subsec:Lipatov-QCD}, we discussed the building blocks of $2\rightarrow N$
scattering in QCD in Regge asymptotics. These are the Lipatov vertex which appears in the 
$2\rightarrow 3$ amplitude and the reggeized gluon propagator that arises from iterating the IR divergent pieces of the virtual contributions to the $2\rightarrow 2$ amplitude. 
In \cite{Lipatov:1982it,Lipatov:1982vv}, Lipatov demonstrated that the corresponding $2\rightarrow N$ amplitude in gravity can be constructed analogously with the building blocks being the gravitational Lipatov vertex and the reggeized graviton propagator. As we shall discuss, the former has a contribution corresponding to the double copy of the QCD Lipatov vertex and the latter is similar to the QCD reggeized gluon propagator with the intercept of the Regge trajectory at $2$ instead of its value of unity in the QCD case. 

Since in Section~\ref{subsec:Lipatov-QCD} we showed that the BFKL ladder and generalizations thereof could be reproduced in a powerful semi-classical RG framework, it is natural to ask whether the same semi-classical approach can be applied to describe 
$2\rightarrow N$ scattering in gravity for $N\gg 1$. From the qualitative arguments 
presented in \cite{Dvali:2021ooc}, and from the derivation we will present in 
Section~\ref{sec:gr-shockwaves} and the subsequent discussion in Section~\ref{sec:Outlook}, we conjecture this to be the case. Establishing it fully will however take considerable work beyond the scope of this paper. 

The conjectured map of the semi-classical CGC EFT to a semi-classical Regge EFT of gravity is by no means obvious since we are dealing with two very different theories. QCD becomes a strongly interacting confining theory in the infrared (or at large impact parameters in the context of the discussion here) while gravity becomes a strongly interacting theory in the  UV, whose structure is that of a yet undermined theory of quantum gravity. More specifically, the Regge limit of $2\to 2$ scattering in gravity, unlike QCD, 
involves several dimensionful scales. Setting $\hbar =1$, the Planck mass ($M_p$) or Planck length ($\ell_p$) is related to the parameter ($\kappa$) or Newton's constant ($G$), 
\be
\frac{\kappa^2}{8\pi}= G = \frac{1}{M_p^2}=\ell_p^2\,,
\ee
and the trans-Planckian scattering regime is specified by taking the center of mass energy $\sqrt{s}\gg M_p$ where we have the hierarchy of scales $ \ell_p\ll R_S \ll b$. Here $b$ is the impact parameter which is conjugate to the momentum transfer $Q$. The  latter combines with $M_p$ to give the dimensionless gravitational coupling
\be
\label{effectiveCoupling}
\alpha(Q) = \frac{Q^2}{M_p^2}~.
\ee 
This is clearly very different from the behavior of $\alpha_s$ in QCD.

The scale $R_S$ is the characteristic Schwarzschild radius set by the center of mass energy $\sqrt{s}$:
\be
R_S \equiv G\sqrt{s} =\frac{\sqrt{s}}{M_p^2}~.
\ee
For incoming particles with impact parameter within $R_S$, classical arguments suggest that a black hole will be formed \cite{Thorne:1972ji}. To avoid encountering complications of near horizon effects, we will restrict ourselves to the regime where the impact parameter $b$ is much larger than the Schwarzschild radius $b\gg R_S$. An specific goal will be to understand if $R_S$ (which in a black hole Quantum Portrait framework~\cite{Dvali:2011aa} is the inverse of the saturation scale $Q_S$) can similarly be extracted from the RG evolution of the $2\rightarrow N$ amplitude to smaller impact parameters. 

\begin{figure}[h]
\centering
\begin{subfigure}[b]{\textwidth}
\centering
\raisebox{-27pt}{\includegraphics[scale=1]{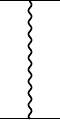} }
+
\raisebox{-27pt}{\includegraphics[scale=1]{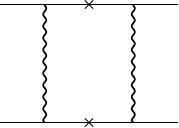} }
+
\raisebox{-27pt}{\includegraphics[scale=1]{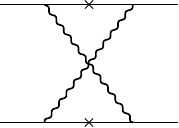} }
+
\raisebox{-27pt}{\includegraphics[scale=1]{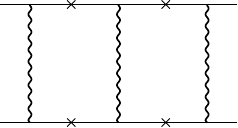} }
+~$\cdots$
\caption{The eikonal scattering series comprised of horizontal ladder and crossed ladder diagrams. The crosses  denote that the propagators of the high energy lines are approximated as $1/(p-k)^2\sim -1/(2p\cdot k)$.} \label{fig:221}
\end{subfigure}
\begin{subfigure}[b]{\textwidth}
\centering
\raisebox{-27pt}{\includegraphics[scale=1]{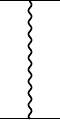} }
+
\raisebox{-27pt}{\includegraphics[scale=1]{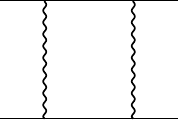} }
+
\raisebox{-27pt}{\includegraphics[scale=1]{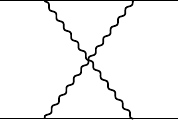} }
+~$\cdots$
\caption{Multiple t-channel graviton exchanges (without the eikonal approximation) where the loops gives rise to the relevant large logarithms shown in Eq.~\eqref{one-loop-graviton}.} \label{fig:222}
\end{subfigure}
\caption{Multiple t-channel graviton exchange diagrams including both eikonal and off-shell propagators. 
}
\end{figure}

It is well known that at large impact parameters $b\gg R_S$, the $2\to 2$ gravitational amplitude eikonalizes in the Regge limit \cite{Amati:1987uf, Muzinich:1987in, Kabat:1992tb}. The diagrams that contribute to eikonalization comes from resumming the horizontal ladder and cross ladder series shown in Fig. \ref{fig:221}. In these diagrams, the eikonal approximation requires that the momenta of the exchanged gravitons are  neglected with respect to those of the high energy lines, namely, $(p_1-k)^2 \approx -2p_1\cdot k$. This replacement is illustrated by the crosses in Fig. \ref{fig:221}. The resummation of this series generates the eikonal amplitude:
\be
\label{EikAmp}
i\mathcal{M}_{\rm Eik}=2 s \int d^{2} \bsb~ e^{-i \mathbf{q} \cdot \mathbf{b}}\left(e^{i \chi(\bsb, s)}-1\right)~,
\ee
where the IR divergent {\it eikonal phase} $\chi(\bsb,s)$ is given by
\be
\chi(\bsb,s)=\frac{\kappa^2 s}{2} \int \frac{d^2 \bsk}{(2 \pi)^2} \frac{1}{\bsk^2} e^{i \bsb \cdot \bsk}~.
\ee

The full one-loop four point amplitude contains the two sets of contributions shown below\footnote{This expression only includes the leading terms in the Regge limit that have an IR divergence. For the complete expression, see for instance Eqs.~(14) and (18) in \cite{Bartels:2012ra}, and references therein. }:
\begin{align}
\label{one-loop-graviton}
    \mathcal{M}^{(1)} \sim \frac{\kappa^2}{8\pi^2}\( - i\pi  s \log\(\frac{-t}{\Lambda^2}\) + t \log\(\frac{s}{-t}\)\log\(\frac{-t}{\Lambda^2}\)\)~\,,
\end{align}
where $\Lambda$ is an IR cutoff. The first term in the parenthesis is from the aforementioned eikonal amplitude and the other is the contribution from off-shell propagators. Since both terms have exactly the same IR divergence in the Regge limit the first term dominates over the second term; the latter is subleading as $-t/s \sim R_S^2/b^2$ and therefore not relevant for the large impact parameter regime $b\gg R_S$ \cite{Giddings:2011xs}. This is qualitatively different from the perturbative QCD case where the double logs dominate over the eikonal phase contribution. For a nice discussion of this difference from the double copy perspective, see \cite{Melville:2013qca}. 

When one takes all-loop iterations\footnote{Einstein gravity is non-renormalizable in the sense of a conventional quantum field theory. When we discuss all-loop contributions, we refer here to contributions that dominate at leading log in the Regge limit that are iterations of the one loop four-point graviton amplitude. Pure gravity in four dimensions is known to be renormalizable to this order~\cite{tHooft:1974toh}; for a nice discussion of UV divergences at two loops, and relevant references, see \cite{Bern:2017puu}. Whether reggeization holds beyond one loop order in gravity is not known. From the modern EFT perspective, renormalizability may not be relevant to this issue since our focus is on the IR regime of the theory~\cite{Donoghue:2022eay}; indeed, it has been argued that black holes UV complete gravity~\cite{Dvali:2010jz}. Though we will not address this issue here, we note that an interesting program to address aspects of this issued would be to consider renormalization in the strong shockwave background. In QCD this does not add anything to the discussion but the situation may be qualitatively different in gravity for the aforementioned reason of UV completion.} of the $2\rightarrow 2$ amplitude, the first term corresponds to multiple scattering contributions which can be resummed into an exponential form (in impact parameter space) and results in the eikonal amplitude in Eq.~\eqref{EikAmp}. The behavior of the iteration of the second term in the parenthesis in Eq.~\eqref{one-loop-graviton} is familiar from the QCD discussion. These are the doubly logarithmic divergent Sudakov logs whose resummation (from multiple graviton exchange) leads to reggeization. Our discussion of the one loop result would suggest that reggeization is kinematically suppressed in gravity as $R_S^2/b^2$ and therefore not important at large impact parameters. 

However as evident from the expression, these double logs are important at smaller impact parameters. Furthermore, they are essential to the problem of interest here - the construction of the $2\rightarrow N$ inelastic amplitude. This is because the IR divergences from the loop terms cancel those from the real emission amplitude in the scattering cross-section for this process. This cancellation is identical to the QCD case and therefore important for the same reason as for particle production in perturbative QCD (and in QED). Thus reggeization should go through in the same manner as in QCD and along with the gravitational Lipatov vertex provide the building blocks for the construction of the 
$2\rightarrow N$ amplitude to all orders to leading logarithmic accuracy. 

In \cite{Lipatov:1982it,Lipatov:1982vv}, Lipatov computed the graviton Regge trajectory\footnote{See \cite{Grisaru:1975tb} for earlier work on graviton reggeization in pure Einstein gravity.}
\be
\label{gravitonReggeTrajectory}
\alpha(t) =-\kappa^2 t  \int \frac{d^2 \bsk}{(2\pi)^2}\frac{1}{\bsk^2\left(\bsq-\bsk\right)^2}\left[\left(\bsk\cdot(\bsq-\bsk)\right)^2\left(\frac{1}{\bsk^2}+\frac{1}{\left(\bsq-\bsk\right)^2}\right)-\bsq^2\right]~,\qquad \bsq^2 = -t
\ee
that gives rise to reggeized graviton propagators. 
As in QCD, the graviton Regge trajectory contributes to the virtual part of the gravitational BKFL kernel as shown in \cite{Lipatov:1982it, Lipatov:1988ii}. 

The real part of the gravitational BFKL kernel gets a contribution from the square of gravitational Lipatov vertex\footnote{This effective vertex capturing the emission of soft gravitons in MRK kinematics was also discovered in the context of closed string scattering \cite{,Ademollo:1989ag,Ademollo:1990sd} and later on in \cite{SabioVera:2011wy, SabioVera:2012zky, Johansson:2013nsa} where double copy relations to the gauge theory Lipatov vertex was made.}. In covariant form, it can be expressed as a double copy of the gauge theory Lipatov vertex $C^\mu$ as 
\be\label{GLV}
\Gamma_{\mu \nu}(\bsq_1, \bsq_2)\equiv \frac12 C_{\mu}(\bsq_1,\bsq_2) C_{\nu}(\bsq_1,\bsq_2) -\frac12 N_{\mu}(\bsq_1,\bsq_2) N_{\nu}(\bsq_1,\bsq_2)~\,,
\ee
where one has an additional double copy of the quantity $N_\m$. This is the soft photon vertex \cite{Weinberg:1965nx}
\be\label{QEDB}
N_\m (\bsq_1,\bsq_2)=\sqrt{\bsq_1^2 \bsq_2^2}~\(\frac{p_{1\m}}{p_1\cdot k}-\frac{p_{2\m}}{p_2\cdot k}\)~,
\ee
dressed by an overall factor $\sqrt{\bsq_1^2 \bsq_2^2}$. These vertices are gauge invariant ($N_\mu k^\mu=\Gamma_{\m\n}k^\n=0$) and traceless ($\eta^{\m\n}\Gamma_{\m\n}=0$). 

An important point worth mentioning about the gravitational Lipatov vertex is that the presence of the $N_\mu N_\nu$ term is required by unitarity. The term with the $C_\m C_\n$ structure alone has a simultaneous pole in the overlapping $s_1 = (k+p_1)^2$ and $s_2 = (k+p_1)^2$ channels. The presence of such a term in an amplitude is forbidden by the so-called Steinmann  relations \cite{Steinmann1,Steinmann2} which ensure that there are no poles and discontinuities in overlapping energy channels. A discussion of these relations in the context of the gravitational Lipatov vertex can be found in \cite{SabioVera:2011wy, SabioVera:2012zky, Johansson:2013nsa,Bartels:2008ce}.

\begin{figure}[h]
\centering
\includegraphics[scale=1]{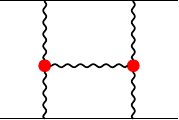}
\caption{The leading corrections to the eikonal scattering series captured by H-diagrams. The red dots indicate gravitational Lipatov vertices.} \label{H-diagram}
\end{figure}

The full content of the $2\rightarrow 3$ ``block" of the $2\rightarrow N$ amplitude in multi-Regge kinematics is represented by the so-called H-diagram~\cite{Amati:2007ak} shown in Fig. \ref{H-diagram}. Albeit subleading as $R_S^2/b^2$ in the eikonal expansion, this provides the leading contribution to inelastic graviton production which becomes increasingly important as $b\rightarrow R_S$. 

For the purposes of the discussion in the following section, it will be useful to recast the covariant formula for the Lipatov vertex into corresponding expressions in light cone gauge where they simplify considerably. Let $\ep_\m(k)$ be the gluon polarization vector and $\ep_{\m\n}(k)$ be the graviton polarization tensor such that (suppressing helicity labels)
\be
\ep_{\m\n}(k)= \ep_\m(k) \ep_\n(k)~,~~~~\ep_\m(k)k^\m=0~.
\ee
In the light-cone gauge $\varepsilon_{+\mu}=0$ one finds that the soft photon emmision factor in Eq.~\eqref{QEDB} takes the form
\begin{align}
\label{QCDLVLC2}
\ep^*_{\m}(k) N^\m(q_1,q_2) = -2\sqrt{\bsq_{1}^2\bsq_{2}^2} \frac{\ep^*_i k_i}{\bsk^2} \equiv \ep^*_i N_i(\bsq_1,\bsq_2)~.
\end{align}
Using this result along with Eq.~\eqref{QCDLVLC1} and the gauge invariance condition $\Gamma_{\m\n}k^\n=0$, one can obtain all nonvanishing components of the gravitational Lipatov vertex in the light cone gauge. The explicit expressions for these are as follows:
\begin{align}
\label{LCLV1}
\Gamma_{ij}(\bsq_1,\bsq_2)&=2\(q_{2i}-k_i\frac{\bsq_{2 }^{2}}{\bsk^{2}}\) \(q_{2j}-k_j\frac{\bsq_{2 }^{2}}{\bsk^{2}}\) - 2k_ik_j\frac{\bsq_{1}^2\bsq_{2}^2}{\bsk^4}\,,\\
\label{LCLV2}
\Gamma_{-i}(\bsq_1,\bsq_2)&=\frac{4k_-}{\bsk_\perp^2}\bigg[ (\boldsymbol{q}_1\cdot \boldsymbol{q}_2) \(q_{2i}-k_i\frac{\bsq_{2 }^{2}}{\bsk^{2}}\) -k_i\frac{\bsq_{1}^2\bsq_{2}^2}{\bsk^2}\bigg]\,,\\
\label{LCLV3}
\Gamma_{--}(\bsq_1,\bsq_2)&=\frac{8k_-^2}{\bsk_\perp^4}\bigg[(\boldsymbol{q}_1\cdot \boldsymbol{q}_2)^2 -\bsq_{1}^2\bsq_{2}^2\bigg]~.
\end{align}

In the next section, we will present an alternative derivation of the central emission Lipatov vertex in a semi-classical framework of Einstein-Hilbert gravity that's exactly analogous to the discussion in Section~\ref{subsec:Lipatov-QCD}. The large occupancy of gravitons resulting from their bremsstrahlung across a wide range of rapidities higher than the rapidity of interest allows one to to treat their dynamics as a static mass distribution that couples coherently to the emitted graviton analogously to the classical color density in the QCD case. 

What we have discussed thus far is analogous to the discussion of the dilute-dilute regime in QCD whose RG evolution is described by the BFKL equation. One may therefore ask whether in gravity one has an identical  classification of inclusive distributions as ``dilute-dilute", ``dilute-dense" and ``dense-dense" as described for the QCD case in Section~\ref{subsec:Lipatov-QCD}. Our discussion in this paper will focus on the simplest dilute-dilute case but we will briefly address the dilute-dense case in Section~\ref{sec:Outlook} in the context of shockwave propagators necessary to promote the $2\rightarrow 3$ computation to the RG description of $2\rightarrow N$ scattering. The explicit computation of these propagators will cast further quantitative light on the gravitational dilute-dense and dense-dense regimes.

Even though the quantitative tools to address the dilute-dense and dense-dense regimes are not fully developed (unlike the QCD case), we believe nevertheless that such regimes must exist on physical grounds. The discussion can be framed in the context of the extensive work of ACV classifying the different regimes that contribute to trans-Planckian scattering. As discussed by ACV, the resummation of all possible contributions due to  graviton exchanges to the $2\to 2$ S-matrix can be expressed generally as an exponentiated form in impact parameter space as $e^{2i\delta(b,E)}$. The phase $\delta(b,E)$ is generically complex with the imaginary part corresponding to inelastic final states describing $2\rightarrow N$ emissions. It can be organized in pure gravity\footnote{In other extended theories of gravity such as string theory there can appear ratios of other length scales for instance $\lambda_s^2/b^2$.} as a power series expansion in $R_S^2/b^2$ and $\ell_p^2/b^2$. Writing the S-matrix in ACV's convention as $S= e^{2i(\delta_0 + \delta_1 + \delta_2 + \cdots)}$, the known phase factors are \cite{Amati:1990xe}:
\begin{align}
    \delta_0 = Gs \log\(\frac{L}{b}\)~,\qquad \delta_1 =\frac{6G^2 s }{\pi b^2} \log s~,\qquad \delta_2 =\frac{2G^3s^2}{b^2}\left[1+\frac{i}{\pi} \log s\left(\log \frac{L^2}{b^2}+2\right)\right] .
\end{align}
Here $\delta_0$ corresponds to the Eikonal phase $\chi(\bsb,s)$ we discussed earlier. The second term $\delta_1$ corresponds to a pure quantum gravity correction $G^2 s/b^2 \sim G s\(\ell_p^2/b^2\)$ which can be ignored for the regime $b\gg R_S\gg \ell_p$. Like $\delta_0$, this contribution is purely real. The final term in ACV's decomposition $\delta_2$ is a classical correction  proportional to $G^3s^2/b^2 \sim Gs \(R_S^2/b^2\)$. One observes that while these terms are weighted by various powers of $R_S^2/b^2$ or $\ell_p^2/b^2$, they all appear with a factor of the dimensionless combination $Gs\gg 1$. One the face of it, this demonstrates that it is inconsistent to truncate the expansion of the exponential $e^{2i\delta(b,E)}$ to a finite number of terms. 

Our interest  is in the inelastic contributions to the $2\rightarrow 2$ S-matrix that correspond to $2\rightarrow N$ final states. In particular, one may ask whether the (unsuppressed) all-order soft rescattering contributions to the {\it hard} $2\rightarrow 3$ inclusive distribution arising from multiple rescattering exchanges be factorized into the Weinberg soft factor containing all soft real and virtual exchanges. This question was addressed both by Lipatov and ACV who both distinguished between soft ``Weinberg" momentum $k\ll q_1$, $q_2$ and the opposite ``Lipatov" semi-hard regime $k\gg q_1$, $q_2$ (where $q_1$, $q_2$ are the $t$-channel  momentum transfers in the multi-Regge $2\rightarrow N$ amplitude and $k$ is the momentum of the emitted graviton), in the terminology of \cite{Ciafaloni:2018uwe}. As already shown by Lipatov, and confirmed by ACV, all Weinberg-type multiple exchanges and radiation can be absorbed into the overall Weinberg soft factor multiplying the semi-hard multi-Regge amplitude. 

On the basis of this discussion, we would conclude that there is indeed a dilute-dilute Lipatov regime of evolution of the semi-hard inelastic $2\rightarrow N$ amplitude in the Regge regime. One may further ask whether the dilute-dilute evolution regime is larger or smaller than it is in QCD before high parton density effects become important. On the one hand, the occupancy $N$ (at a given impact paramter) of inelastically produced {\it hard} final state gravitons grows much more rapidly than the multiplicity of gluons in QCD due to the larger Regge trajectory in the former. On the other hand, these occupancies have to be much larger in gravity for semi-hard graviton rescattering and recombination to be significant. Regardless of the width in rapidity of the dilute-dilute regime,  the growth of graviton occupancy must saturate as it does in QCD because the probability for such emissions at a fixed impact parameter cannot exceed unity. It has been argued on information theoretic grounds that the growth saturates for $N=1/\alpha_{\rm crit}$, where $\alpha_{\rm crit}=\alpha(Q_s)\equiv Q_S^2/M_p^2$ and $Q_S=1/R_S$ is an emergent scale~\cite{Dvali:2010jz,Dvali:2011aa,Dvali:2020wqi}. As noted earlier, the similarities between this dynamics and that of the CGC has been discussed previously~\cite{Dvali:2021ooc}. We will return to the quantitative realization of this picture {\it a la} CGC in Section~\ref{sec:dilutedense}.


\section{Semi-classical scattering of gravitational shockwaves}
\label{sec:gr-shockwaves}

\subsection{Gravitational shockwaves}
\label{sec:2.1}
Aichelburg and Sexl \cite{Aichelburg:1970dh} showed that when a Schwarzschild black hole characterized by mass $m_H$ is given an infinite boost $\gamma \to \infty$ (say along the positive $z$ direction) then in the limit $m_H\to 0$, with total energy $\mu_H= \gamma\, m_H $ held fixed, one obtains the shockwave spacetime
\begin{align}
\label{ASmetric}
ds^2 = &~ 2\,dx^+dx^- -\delta_{ij}dx^i dx^j + 8\,\mu_H \,G \,\delta(x^-)\log(\Lambda |\bsx|) \(dx^-\)^2~.
\end{align}
Here $\Lambda$ is an IR cutoff scale. This metric is a solution to Einstein's equation (Eq.~\eqref{EinsteinEq} in Appendix~\ref{appB}) with a nonvanishing energy-momentum (EM) tensor given by $T_{\m\n} = \delta_{\mu-}\delta_{\nu-} \delta(x^-) \delta^{(2)}(\bsx)$ required to support this geometry. This is the EM tensor of a massless point particle located at the origin of the transverse space $\bsx=0$. 

Motivated by our discussion in Section \ref{subsec:Lipatov-QCD}, one can generalize this form of the EM tensor to include a source with transverse spatial density $\rho_H(\bsx)$ which has the shockwave profile\footnote{A subtle point is that for a Schwarzschild black hole at rest the energy-momentum tensor vanishes. When boosted to $\gamma\to\infty$, a nonvanishing stress-tensor develops as a consequence of the singular nature of the limit which erases the information of the horizon. However it is important to keep in mind that the mass distributions we will discuss are not to be thought of as strict 
$\delta$-functions in $x^-$. As in Section \ref{subsec:Lipatov-QCD}, this source in Regge asymptotics is the ``static" distribution of gravitons at higher rapidities; there is therefore an implicit rapidity scale dependence of this distribution.}
\begin{align}
\label{EMtensor1}
T_{\m\n} = \delta_{\mu-}\delta_{\nu-} \mu_H \delta(x^-) \rho_H(\bsx)~.
\end{align}
The resulting more general shockwave spacetime has the metric
\be
\label{denseBgnd1}
ds^2 = 2dx^+dx^- -\delta_{ij}dx^i dx^j + f(x^-,\bsx)\(dx^-\)^2 ~,
\ee
where 
\begin{align}
\label{backgroundg}
f(x^-,\bsx) &= 2\kappa^2\mu_H \delta(x^-)\frac{\rho_H(\bsx)}{\square_\perp} = \frac{\kappa^2}{\pi}\mu_H \delta(x^-) \int d^2\bsy ~\ln\Lambda |\bsx-\bsy| \rho_H(\bsy)~.
\end{align}
We used in the second equality the Green's function of the two-dimensional Laplacian $\square_\perp\equiv\delta_{ij}\p_i\p_j$. In this singular form, where the source $\rho_H$ appears linearly and a delta function appears in the metric, one sees that the spacetime is flat in the regions in front ($x^-<0$)  and  behind ($x^->0$) the shockwave. However the inertial frames in these regions are not identical. They are related by a coordinate transformation of the Minkowski vacuum. This is to be expected intuitively since the passing shock should affect spacetime measurements differently in these regions \cite{Dray:1984ha,tHooft:1996rdg}. To see it more rigorously, we transform to the $y$-coordinate frame which is related to the $x$-coordinate by the discontinuous transformation 
\begin{align}
\label{xytransformation}
\begin{split}
x^- =& ~ y^-,\qquad x^i = ~ y^i - \kappa^2\m_H y^- \Theta(y^-)\frac{\p_i}{\square_\perp} \rho_H(\bsy) ~,\\[5pt]
x^+ =& ~ y^+ - \kappa^2\m_H \Theta(y^-) \frac{\rho_H(\bsy)}{\square_\perp} + \frac12  \kappa^4 \m_H^2~y^-\Theta(y^-) \(\frac{\p_i}{\square_\perp} \rho_H(\bsy)\)^2~.
\end{split}
\end{align}
This transformation comes from analyzing null geodesics passing through the spacetime specified in Eq.~\eqref{denseBgnd1}, a point we will return to shortly. The metric in the $y$-coordinate system takes the form
\begin{align}
\label{denseBgnd2}
    ds^2 = 2dy^+dy^- -g_{ij}dy^idy^j~,
\end{align}
where $g_{ij}$ has a nonlinear dependence on the source $\rho_H$ and is given by
\begin{align}
    g_{ij} = &~ \delta_{ij}-y^-\Theta(y^-)\bigg[2\kappa^2\m_H~\frac{\p_i\p_j}{\square_\perp}\rho_H(\bsy)- \kappa^4 \m_H^2~y^-\(\frac{\p_i\p_k}{\square_\perp}\rho_H(\bsy)\) \(\frac{\p_j\p_k }{\square_\perp}\rho_H(\bsy)\)  \bigg]~.
\end{align}
The continuous form of the metric in Eq.~\eqref{denseBgnd2} is an exact solution of Einstein's equations with the EM tensor in Eq.~\eqref{EMtensor1}. It makes manifest that the region in front of the shock ($y^-<0$) is the Minkowski vacuum while the region after the shock ($y^->0$) is a pure gauge transformation of the Minkowski vacuum. (The latter is seen by computing the Riemann tensor $R_{\m\n\rho\sigma}$  of the metric in the $y^->0$ region which turns out to vanish even though its connection coefficients do not vanish.) 

This is exactly analogous to the Yang-Mills shockwave solution in the Color Glass Condensate EFT for which the field strength tensor vanishes before and after the shockwave even though the gauge fields do not vanish; they are distinct pure gauge solutions separated by the gluon shockwave~\cite{McLerran:1993ka,McLerran:1993ni}.


\subsection{Linearized fluctuation around the shockwave background and the gravitational Wilson line}
\label{sec:3b}
Having discussed the properties of gravitational shockwave metric, we will now analyze the structure of small fluctuations around it. At the classical level, small fluctuations are governed by linearized Einstein's equations around the shockwave background. For this purpose, it is simpler use the singular form of gravitational shockwave given in Eq.~\eqref{denseBgnd1}. We start by writing small perturbations about the metric as 
\begin{align}
    g_{\m\n} = \bar{g}_{\m\n}+\kappa\, h_{\m\n}~.
\end{align}
We weighted the fluctuation field $h_{\m\n}$ here by the coupling $\kappa$ so that its kinetic term is canonically normalized. Working in the light cone gauge
\be
h_{\mu+}=0~,
\ee
the linearization procedure of Einstein's equations results in the following set of second order equations:
\begin{align}
\begin{split}
\label{sourcelessEOM}
    &\square h_{ij} - \bar{g}_{--}\p_+^2 h_{ij} = 0~, \qquad \square h_{i-} - \bar{g}_{--}\p_+^2 h_{i-} = \p_+ h_{ij}\p_j \bar{g}_{--}~,\\[5pt]
    &\square h_{--} - \bar{g}_{--}\p_+^2 h_{--} = \(\p_i\p_j \bar{g}_{--}\) h_{ij}  + 2 \(\p_i\bar{g}_{--}\) \p_j h_{ij} ~.
\end{split}
\end{align}
These equations are not all independent since there are first order constraint relations among various components of $h_{\m\n}$: $\p_+h_{-i}=\p_jh_{ij}~, \p_+ h_{--} =\p_i h_{-i}$. Furthermore Einstein's equations set $h\equiv \delta_{ij}h_{ij}=0$. These equations imply that $h_{ij}$ are the independent components of the metric fluctuations corresponding to the physical degrees of freedom.

Next, we solve these equations. In the vicinity of $x^-=0$, the transverse derivatives acting on $h_{ij}$ in Eq.~\eqref{sourcelessEOM} can be neglected and we obtain 
\be
\p_-h_{ij}-\frac12\bar{g}_{--}\p_+ h_{ij} =0~,
\ee
which can be solved for the fluctuation at $x^-=x^-_0$. The solution is given by
\be
\label{solij}
h_{ij}(x^+,x^-,\boldsymbol{x}) = V(x^-,\boldsymbol{x}) h_{ij}(x^+,x^-=x^-_0,\boldsymbol{x})\,,
\ee 
where the gravitational Wilson line operator $V$ is given by
\be\label{VWilson}
V(x^-, \boldsymbol{x}) \equiv \exp \(\frac12 \int_{x^-_0}^{x^-} dz^- \bar{g}_{--}(z^-, \boldsymbol{x})\,\p_+\)\,.
\ee
Using the constraint relations, the solution for the other two components can be written as
\begin{align}\label{solim}
h_{-i}(x^-) &= V(x^-)h_{-i}(x_0^-) +\(\p_j V\) \frac{1}{\partial_+} h_{ij}(x_0^-)~,\\[5pt]
\label{solmm}
h_{--}(x^-) &= V(x^-)h_{--}(x_0^-) + 2 \(\p_i V\) \frac{1}{\partial_+} h_{-i}(x_0^-)+\(\p_i\p_j V\) \frac{1}{\partial_+^2} h_{ij}(x_0^-)~.
\end{align}
One can verify that the above result satisfies the equations of motion in Eq.~\eqref{sourcelessEOM}. The solutions in Eqs.~\eqref{solij}, \eqref{solim}, \eqref{solmm} are gluing formulae that connect plane wave evolution from one side of the shockwave to plane wave evolution on the other side. The gravitational Wilson line operator $V$ appearing in these formulas are shift operators that act along the shockwave and whose magnitude is a function of the energy of the shockwave and its transverse distribution. 

The exact analog of Eqs.~\eqref{solij}, \eqref{solim}, \eqref{solmm} in the gauge theory case was worked out in \cite{Gelis:2005pt}. These results have a flavor of a double copy relation. In particular, the gauge theory Wilson line and the gravitational Wilson line are related by the color-kinematic replacement rule which has been extensively discussed in the literature \cite{Saotome:2012vy, Melville:2013qca}; we will briefly revisit it in section \ref{sec:IIID}.


\subsection{Shockwave collisions}

In this subsection, we will address the problem of the collision of two gravitational shockwaves in the dilute-dilute approximation analogous to the QCD case discussed in section~\ref{subsec:Lipatov-QCD}. In our setup, we consider two incoming gravitational shockwaves along the $z$ axis separated by impact parameter $b$ in the transverse plane with the collision point at $z=t=0$. These shockwaves are generated by the EM tensor (in the region $t<0$),
\begin{align}
\begin{split}
\label{EMtensor2}
T_{\m\n} = &~ \delta_{\mu-}\delta_{\nu-} \m_H \,\delta(x^-) \rho_H(\bsx) +\delta_{\mu+}\delta_{\nu+} \m_L \delta(x^+) \rho_L(\bsx)~.
\end{split}
\end{align}

\begin{figure}[ht]
\centering
\includegraphics[scale=1]{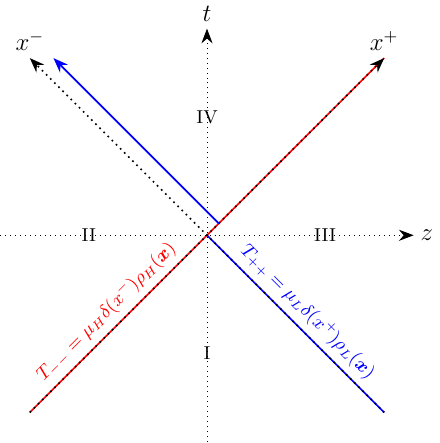}
\caption{Trajectories of colliding ultrarelativistic particles. The dilute-dense approximation is the one where the trajectory of the ``heavy" graviton shockwave (depicted in red) is unchanged whereas that of the light shockwave (depicted in blue) undergoes a shift (according to Eq.~\eqref{geodesicSol}). The trajectories in the future of the collision are bent slightly inwards but this effect is small at large impact parameters between the two shocks in the transverse plane.}
    \label{spacetime1}
\end{figure}

This is depicted in Fig.~\ref{spacetime1} where II and III are space-like regions of respectively H and L shockwaves before the collision that correspond to the respective coordinate transformation of Minkowski vacua (in other words they are pure gauges -- see discussion in Section \ref{sec:2.1} above) while region I is the common Minkowski vacuum shared by both the shockwaves. Finally, region IV corresponds to the future of collision in which the EM tensor of each of the shocks will get modified and  backreact to create a radiative spacetime with a nonvanishing curvature in region IV. 

We will first calculate the correction to the EM tensor and the modified metric in region IV in the dilute-dilute approximation where we keep terms to linear order in $\rho_H$ and $\rho_L$. (In contrast, the dilute-dense approximation corresponds to keeping terms which are all order in $\rho_H$ but only linear order in $\rho_L$.)  We will then set up the equations of motion of the metric in the dilute-dense approximation. However for our purposes, it suffices to solve for the modification of EM tensor and the metric in the dilute-dilute approximation since we are working in the regime $b>R_S$. (See the discussion in Section \ref{sec:intro}.)

\subsubsection{Equations of motion}

Treating the spacetime created by shockwave H as background, we consider small perturbations $h_{\m\n}$ around it:
\begin{align}
\label{perturb}
g_{\m\n} = \bar{g}_{\m\n} + h_{\m\n}~.
\end{align}

Here $\bar{g}_{\m\n}$ is the background metric tensor appearing in Eq.~\eqref{denseBgnd1}. We decompose the perturbation $h_{\m\n}$ into a term $h_{\mu\nu}^{(1)}$ of order $O(\rho_L)$ which is sourced by $T_{++}$ (appearing in Eq.~\eqref{EMtensor2}) and a term $h_{\mu\nu}^{(2)}$ of order $O(\rho_H\rho_L)$ coming from the backreaction of the corrected EM tensor at this order. Therefore
\begin{align}
    h_{\mu\nu} = h_{\mu\nu}^{(1)}+\kappa \,h_{\mu\nu}^{(2)}~.
\end{align}
We weighted $h_{\mu\nu}^{(2)}$ by a factor of $\kappa$ in order to ensure the corresponding kinetic term is canonically normalized. Working in the light cone gauge\footnote{This gauge condition leaves a residual gauge freedom unfixed. We will fix the gauge completely by demanding transversality of physical polarizations of the graviton: $\p^\m h_{\m\n}=0$.} $h_{\mu+}=0$,  linearizing Einstein's field equations around $\bar{g}_{\m\n}$, and further processing the results in the equation of motion for the traceless field $\tilde{h}_{ij}$,
\begin{align}
\label{metricEq}
&\bar{g}_{--}\p_+^2 \tilde h_{ij}- \square \tilde h_{ij}= \kappa^2\bigg[\(2\p_i\p_j-\square_\perp \delta_{ij}\)\frac{1}{\p_+^2}T_{++}  +2T_{ij}-\delta_{ij}T-\frac{2}{\p_+} \(\p_iT_{+j}+\p_jT_{+i}-\delta_{ij}\p_k T_{+k}\) \bigg]~.
\end{align}
Here $\tilde{h}_{ij} \equiv h_{ij}-\frac12 \delta_{ij} h$ where $h=\delta_{ij}h_{ij}$; see Appendix \ref{appB} for details. Further, $\square$ is the d'Alembertian operator and $T\equiv \delta_{ij}T_{ij}$. 

The other components of the metric fluctuations are obtained from the solution for $\tilde{h}_{ij}$ using the constraint relations
\begin{align}
\label{constraintrel}
\p_+h_{-i} =\p_j\tilde h_{ij}+\kappa^2\[\frac{2}{\p_+}T_{+i} - \frac{\p_i}{\p_+^2} T_{++}\]~\,; \,\, \p_+^2 h_{--} = \p_i\p_j\tilde h_{ij} -\kappa^2\[\frac{\square_\perp}{\p_+^2} T_{++}  - T  - 2T_{+-} +\bar{g}_{--} T_{++}\]~\,.
\end{align}
We refer the reader to Appendix \ref{appB} for further details of these results. These equations are to be further supplemented with the covariant conservation equations of the EM tensor which are given in Eq.~\eqref{EMConservation}. However the conservation laws by themselves do not uniquely determine the evolution of EM tensor. This is evident from Eq.~\eqref{EMConservation} since there are more unknowns than the number of equations. 
As we will now elaborate, to address this problem we will need to consider the geodesic motion~\cite{Taliotis:2010pi, Constantinou:2013tia, Goldberger_2017}  of the ultrarelativistic distribution of particles $\rho_L$ as they cross the shockwave background of $\rho_H$. 


\subsubsection{Geodesic motion and evolution of EM tensor}
The task at hand is to compute how the EM tensor of the ultrarelativistic distribution of particles L gets modified as they cross the region of influence of gravitational field of shockwave H. We will first consider the point particle approximation of the second shockwave where we take $\rho_L(\bsx)=\delta^{(2)}(\bsx-\bsb)$. The covariant EM tensor of a spinless point particle moving along a worldline $X^\m(s)$ is given by 
\be
\label{PPEMT}
T^{\m\n}(x) = \frac{\mu_L}{\sqrt{-\bar{g}}} \int_{-\infty}^\infty d\l~ \dot{X}^\m \dot{X}^\n~ \delta^{(4)}(x-X(\l))~.
\ee
Here $\bar{g}=-1$ is the determinant of the background metric in Eq.~\eqref{denseBgnd1} and the dot denotes differentiation with respect to  the worldline parameter $\l$. The form of the EM tensor in Eq.~\eqref{PPEMT} follows from the relativistic action of a massless point particle. (See Appendix \ref{appC}.) It is covariantly conserved provided the worldline $X^\m(\l)$ satisfies the geodesic equation
\be
\label{geodesicEq}
\ddot{X}^\m+\Gamma^\mu_{\nu\rho} \dot{X}^\n \dot{X}^\rho=0~,\qquad g_{\nu\rho}\dot{X}^\n\dot{X}^\rho=0~.
\ee
The second relation ensures that the geodesic is null. 

Hence in order to determine corrections to $T_{\m\n}$ it suffices to compute the worldline $X^\m(\l)$ from  Eq.~\eqref{geodesicEq}. The nonvanishing connection coefficients are
\be
\Gamma^{+}_{--}=\frac12 \partial_-\bar{g}_{--}~,~~ \Gamma^{+}_{-i}=\Gamma^{i}_{--}=\frac12 \partial_i\bar{g}_{--}~.
\ee
Upon solving Eq.~\eqref{geodesicEq} with appropriate boundary condition at negative times one finds, 
\begin{align}
\begin{split}
\label{geodesicSol}
    X^- &= \lambda~,~~X^i = b^i -  \kappa^2\mu_H X^-\Theta(X^-) \frac{\p_i\rho_H(\bsb)}{\square_\perp}~\\
    X^+ &= -  \kappa^2\mu_H \Theta(X^-) \frac{\rho_H(\bsb)}{\square_\perp} + \frac{\kappa^4 \mu_H^2}{2}X^- \Theta(X^-) \(\frac{\p_i\rho_H(\bsb)}{\square_\perp}\)^2~,
\end{split}
\end{align}
The null geodesic is continuous along the $x^-$ and transverse directions (as a function of $X^-$) but acquires a discontinuity along the $x^+$ direction after crossing the shockwave H at $X^-=0$, as shown in Fig. \ref{spacetime1}. This is precisely the content of the coordinate transformation in Eq.~\eqref{xytransformation} noted in Section \ref{sec:2.1}. (The transformation in Eq.~\eqref{xytransformation} is a solution to the null geodesic equation with generic boundary conditions.) 

Using Eq.~\eqref{PPEMT}, the result in  Eq.~\eqref{geodesicSol} allows us to reconstruct all components of particle L's EM tensor in the dilute-dense approximation. The detailed expressions are given in Appendix \ref{appC}. In what follows, we will need the formulas in the dilute-dilute approximation for the {\it full} EM tensor; in other words, we need to sum up the changes in EM tensors of both particle H and L. However in light cone gauge, as shown in Appendix \ref{appC}, the EM tensor of particle H does not get any corrections from the gravitational background of particle L. The results for the nonvanishing components of the EM tensor are 
\begin{align}
\label{EMlower}
\begin{split}
T_{++} &= \mu_L\delta(x^+)\rho_L +\kappa^2\mu_H\mu_L \Theta(x^-)\bigg[ \delta'(x^+)\frac{\rho_H}{\square_\perp} \rho_L + x^-\delta(x^+) \p_i\(\frac{\p_i\rho_H}{\square_\perp} \rho_L\)\bigg]~,\qquad T_{--} = \mu_H \delta(x^-)\rho_H~, \\[10pt]
T_{-+} &= \kappa^2\mu_H\mu_L \delta(x^+)\delta(x^-) \frac{\rho_H }{\square_\perp} \rho_L ~,\qquad T_{+i} = \kappa^2\mu_H\mu_L \delta(x^+)\Theta(x^-) \frac{\p_i\rho_H}{\square_\perp} \rho_L~.
\end{split}
\end{align}
This result was a consequence of the strict point particle approximation. However since we are neglecting tidal effects, we can generalize the point particle delta function to a finite transverse source distribution  $\rho_L$, with  details provided in Appendix \ref{appC}. 

An important detail is that in the point particle approximation at finite impact parameter the solution has the freedom of adding a ``contact term" of the form $\rho_H\rho_L$ to the solution of the EM tensor. This gives a $\delta^{(2)}(\bsx)\delta^{(2)}(\bsb-\bsx) = \delta^{(2)}(\bsb)\delta^{(2)}(\bsb-\bsx)$ contribution that vanishes for $|\bsb|>R_S$ which is the approximation for which the point particle computation is valid. However such a contact term in position space gives a finite contribution in momentum space. The coefficient of such contact terms cannot be fixed by the point particle analysis and the freedom to add them will of course affect the final solution. This  freedom should in principle be fixed by other physical considerations; in our case, this would be the unitarity of multi-particle production (discussed below Eq. \eqref{QEDB}). With this in mind in the solution for $T_{++}$ above, we added such a term by hand in the second term in the square brackets. As we will see in the next subsection, this results in the correct expression for the Lipatov emission vertex\footnote{As shown in Appendix \ref{appC}, the geodesic analysis for $T_{++}$ gives 
\be
T_{++} = \mu_L\delta(x^+)\rho_L +\kappa^2\mu_H\mu_L \Theta(x^-)\bigg[ \delta'(x^+)\frac{\rho_H}{\square_\perp} \rho_L + x^-\delta(x^+) \frac{\p_i\rho_H}{\square_\perp} \p_i\rho_L \bigg]~.
\ee
The addition of the contact term is of the form
\be
\kappa^2\mu_H\mu_L x^-\Theta(x^-)\delta(x^+) \rho_H\rho_L~,
\ee
which vanishes in the point particle limit for large impact parameters.
}.


\subsubsection{Solution and the Lipatov vertex}

We first solve for $\tilde{h}_{ij}^{(1)}$, which is the fluctuation in the background created by the light shockwave. In this approximation, Eq. \eqref{metricEq} becomes
\begin{align}
-\square \tilde h_{ij}^{(1)}= \kappa^2\mu_L x^+\Theta(x^+)\(2\p_i\p_j-\square_\perp \delta_{ij}\)\rho_L+O(\kappa^4)~,\no
\end{align}
where on the r.h.s we plugged in the order $O(\rho_L)$ expression for $T_{++}$ from Eq.~\eqref{EMlower}. Note that $\tilde{h}_{ij}^{(1)}$ is independent of $x^-$ since the right side of this equation is independent of $x^-$. Therefore $\square \tilde h_{ij}^{(1)}=-\square_\perp \tilde h_{ij}^{(1)}$, which leads to the result 
\be
\label{htij1}
\tilde h_{ij}^{(1)} = \kappa^2\mu_L x^+\Theta(x^+)\(\frac{2\p_i\p_j}{\square_\perp}-\delta_{ij}\)\rho_L+O(\kappa^4)~.
\ee

At order $O(\rho_H\rho_L)$, the solution for the field $\tilde{h}_{ij}^{(2)}$ is obtained by inserting the solution in Eq.~\eqref{htij1} and the result for the corrected EM tensor in Eq.~\eqref{EMlower}, along with Eq.~\eqref{backgroundg}, into Eq.~\eqref{metricEq}. We then find
\begin{align}
\label{metricSol}
    \square\tilde{h}_{ij}^{(2)} =& 2\delta(x^+)\delta(x^-) \frac{\rho_H}{\square_\perp}P_{ij}\frac{\rho_L}{\square_\perp}-\Theta(x^+)\Theta(x^-)\bigg[P_{ij}\( \frac{\rho_H}{\square_\perp} \rho_L + x^+x^- \p_k\(\frac{\p_k\rho_H}{\square_\perp} \rho_L\)\)\no\\[5pt]
    &-2\bigg\{\p_i \( \frac{\p_j\rho_H}{\square_\perp}\rho_L \)+\p_j \( \frac{\p_i\rho_H}{\square_\perp}\rho_L \)-\delta_{ij} \p_k \( \frac{\p_k\rho_H}{\square_\perp}\rho_L \)\bigg\}\bigg]~,
\end{align}
where $P_{ij}=2\p_i\p_j-\delta_{ij}\square_\perp$. We suppressed in this expression an overall factor of $\kappa^3\mu_H\mu_L$ on the right side for clarity. This equation can be easily integrated in Fourier space, which gives 
\begin{align}
\label{metricSolFt}
    k^2 \tilde{h}_{ij}^{(2)}(k)=&~\int \frac{d^2\bsq_2}{\(2\pi\)^2}\bigg[2P_{ij}(\bsq_2)-\frac{\bsq_2^2}{k_+k_-}\bigg\{P_{ij}(\bsk)\(1+\frac{\bsk\cdot \bsq_1}{k_+k_-}\)-2\(k_i q_{1j}+k_j q_{1i}-\delta_{ij}\bsk\cdot \bsq_1\)\bigg\}\bigg] \frac{\rho_H(\bsq_1)}{\bsq_1^2}\frac{\rho_L(\bsq_2)}{\bsq_2^2}
\end{align}
where we used the shorthand notation $P_{ij}(\bsp) \equiv 2\,p_ip_j-\delta_{ij}\bsp^2$.
The transverse momenta are constrained to satisfy $\bsk=\bsq_1+\bsq_2$.  Additional details in the derivation of this result are provided in Appendix \ref{appD}. 

To extract the Lipatov vertex from this result we need to put the momenta $k$ of the graviton $\tilde{h}_{ij}^{(2)}$ on-shell, $2k_+k_- - \bsk^2=0$, with the Lipatov vertex being the residue of the $1/k^2$ pole. After simple manipulations, and restoring the 
factor $\kappa^3\mu_H\mu_L$,  we find 
\begin{align}
\label{hijfinalresult}
    \tilde{h}_{ij}^{(2)}(k) = \frac{2\kappa^3\mu_H\mu_L}{k^2+i\epsilon k^-} \int \frac{d^2\bsq_2}{\(2\pi\)^2}\, \Gamma_{ij}(\bsq_1,\bsq_2) \frac{\rho_H}{\bsq_1^2}\frac{\rho_L}{\bsq_2^2} ~,
\end{align}
where $\Gamma_{ij}$ is the gravitational Lipatov vertex defined in Eq.~\eqref{LCLV1}. 

As noted in the previous subsection, this result crucially relied on the addition of the contact term in the result of $T_{++}$. Without such a term, we would only reproduce the strict Yang-Mills double copy $C_\m C_\n$ part of the Lipatov vertex correctly. However to get the $N_\m N_\n$ term (which we recall from the discussion in Section \ref{subsec:Lipatov-gravity} is required by unitarity), such a contact term in the solution for $T_{++}$ is necessary. Its presence cannot be argued for strictly on classical grounds and in our semi-classical framework likely comes from a consistent application of Cutkosky's rules in strong gravitational backgrounds. In other words, the point particle approximation is a bad one even at large impact parameters when considering multiparticle production. We will return to this point in future work.

To complete our derivation, one can in a similar manner to the derivation of Eq.~\eqref{hijfinalresult} work 
out the expressions for $h_{-i}$ and $h_{--}$ (with further details in Appendix \ref{appD}):
\begin{align}
\begin{split}
\label{hfinalresult}
    h_{-i}^{(2)}(k) &= \frac{\kappa^3 s}{k^2+i\epsilon k^-} \int \frac{d^2\bsq_2}{\(2\pi\)^2} \Gamma_{-i}(\bsq_1,\bsq_2) \frac{\rho_H}{\bsq_1^2}\frac{\rho_L}{\bsq_2^2} ~, \qquad h_{--}^{(2)}(k) = \frac{\kappa^3 s}{k^2+i\epsilon k^-} \int \frac{d^2\bsq_2}{\(2\pi\)^2} \Gamma_{--}(\bsq_1,\bsq_2) \frac{\rho_H}{\bsq_1^2}\frac{\rho_L}{\bsq_2^2} ~,
\end{split}
\end{align}
where $\Gamma_{-i}$, $\Gamma_{--}$ were given in Eqs.~\eqref{LCLV2} and  \eqref{LCLV3} and $s=2\mu_H\mu_L$ is the center of mass energy squared. 

Eq.~\eqref{hfinalresult} is the principal result of our paper. It demonstrates that using a purely semi-classical approach in precise analogy to the Yang-Mills computations in \cite{Blaizot:2004wu} and \cite{Gelis:2005pt}, one recovers the gravitation Lipatov vertex. Our result is also very suggestive that $\rho_{H}/q_1^2$ and $\rho_L/q_2^2$ have the structure of reggeized graviton propagators. Quantum corrections to the bare propagators can be absorbed in the rapidity scale dependent source distributions thus providing a Wilsonian RG interpretation of reggeization analogous to the QCD interpretation. We will discuss this perspective further in Section~\ref{sec:Outlook}. 


\subsubsection{Relationship between gauge and gravity shockwave collisions}
\label{sec:IIID}
Akhoury and Saotome in \cite{Saotome:2012vy} pointed out a relationship between gauge theory shockwave and gravitational shockwave in terms of a precise double copy relation. The gauge theory shockwave (derived in Appendix \ref{appA}) takes the form
\begin{align}
    A_- = -g\delta(x^-)\frac{\rho_H(\bsx)}{\square_\perp} 
\end{align}
whereas the gravitational shockwave takes the form\footnote{In this formula, we rescaled the expression for $g_{--}$ by a factor of $\kappa$ with respect to the expression given in Eq.~\eqref{denseBgnd1} since this normalization leads to a canonically normalized kinetic term for $g_{--}$ when viewed as fluctuating field. We observe a relative minus sign in the double copy. This is because one is comparing $(-1)^n \mathcal{M}_n$ n-point process in gravity to $(-1)^{n-1} \mathcal{A}_n$ n-point process in gauge theory~\cite{Saotome:2012vy}.} (we repeat the expression here with the canonically normalized gravitational field)
\begin{align}
    -\frac{1}{2}\,g_{--}=-\kappa \delta(x^-)\frac{\mu_H\rho_H(\bsx)}{\square_\perp}~.
\end{align}
One observes that a replacement of the color charge density $\rho_H$ in QCD to the mass density $\mu_H \rho_H$ in gravity 
and $g\to \kappa$ in the gluon shockwave result gives the expression for the gravitational shockwave. 

At the next order in the coupling we have for the gauge theory result\footnote{Here we changed slightly the notation for the color charge densities with respect to the appendix such that $\rho_H \equiv \rho_H^a T^a$ denotes the color charge density matrix and used the identity $i f^{abc}\rho_H^b\rho_L^c =-(T^b)_{ca} \rho_H^b\rho_L^c = \[\(\rho_H \cdot T\)\rho_L\]_a$)} derived in   Appendix \ref{appA}, 
\be
a_{i}(k) = \frac{g^3}{k^2+i\epsilon k^-}\int \frac{d^2\bsq_2}{(2\pi)^2} C_i (\bsq_1,\bsq_2) \frac{\rho_H \cdot T}{\bsq_1^2}\frac{\rho_L}{\bsq_2^2} ~.
\label{eq:QCD-three-vertex}
\ee
As for the replacements above, we recover Eq.~\eqref{hijfinalresult} by replacing 
$g\to \kappa$, the color charge densities $\rho_{H/L}$ with the mass densities $\mu_{H/L}\,\rho_{H/L}$, and the QCD Lipatov vertex $C_i$ with the gravitational Lipatov vertex $\Gamma_{ij}$. As mentioned previously, the connections of the gravitational classical double copy to the Yang-Mills one was observed earlier~\cite{Goldberger:2016iau} but the Lipatov double copy of the emission vertices was not pointed out there. Indeed the particular form of Eq.~\eqref{eq:QCD-three-vertex} is a consequence of ensuring the double copy structure of the Lipatov vertices is manifest ({\it which are both kinematic factors}) rather than the color-kinematic relations. The experience with $2\rightarrow N$ amplitudes in Regge asymptotics suggests that the Lipatov double copy relation is the robust quantity rather than the color-kinematic duality. We expect however that upon taking the ultrarelativistic limit of the emission formulae in \cite{Goldberger:2016iau}, and applying the color-kinematic duality, one should recover the Lipatov double copy. This connection has been made explicit by us in a separate publication~\cite{Raj:2023iqn}.

\section{Outlook}
\label{sec:Outlook}

In the previous section, we established that the Lipatov emission vertex of 
the gravitational $2\rightarrow 3$ radiation amplitude in Regge asymptotics, can be obtained from a semi-classical computation of the first order corrections in the metric produced in collision of two shockwaves. The results are very suggestive that powerful strong field semi-classical methods developed in the Color Glass Condensate EFT of QCD can potentially be employed in gravity (despite the substantial differences in the two theories) due to common universal features of Regge asymptotics. Here we will outline  directions to pursue in extensions of this work. One of these is to compute the spectrum of gravitational radiation using strong field techniques. The other is to extend our dilute-dilute studies to the dilute-dense regime of trans-Planckian gravitational scattering and to develop the renormalization group tools that can help understand  black hole formation as arising from the dynamics of wee gravitons. This latter study will of course influence 
the quest to observe signals of quantum effects in gravitational wave radiation. A final topic of interest is the possible formulation of the commonalities between the overoccupied dynamics of QCD and gravity in the language of asymptotic symmetries. 

\subsection{Gravitational bremsstrahlung}

From the result in Eq.~\eqref{hijfinalresult}, one can calculate the spectrum of energy $E^{\rm GW}$ carried by gravitational wave radiation as 
\be
\label{spectrum}
\frac{dE^{\rm GW}}{d\omega d\Omega} = \frac{1}{2\pi^2}\omega^2 \sum_\lambda |\mathcal{M}^{(\lambda)}|^2~.
\ee
Here $\omega$ and $\Omega$ are, respectively, the Fourier conjugate of $x^+=(t+z)/\sqrt{2}$ and the solid angle. The sum on the r.h.s is over the two physical polarizations $\lambda$ of the gravitational wave amplitude $M^{(\lambda)}$, obtained by contracting $\tilde{h}_{ij}^{(2)}(k)$ computed in Eq.~\eqref{hijfinalresult} with the physical graviton polarization tensors $\varepsilon_{ij}^{(\lambda)}$:
\be
\mathcal{M}^{(\lambda)} = k^2 \tilde{h}_{ij}^{(2)}(k) \varepsilon_{ij}^{(\lambda)}~.
\ee
This follows from the standard LSZ reduction formula; see for instance \cite{Blaizot:2004wu}. We note that the latter reference performs a parallel analysis of the gluon emission amplitude, spectrum and average multiplicity of gluons produced in proton-nucleus collisions in Regge asymptotics using the semi-classical approach where one solves classical Yang-Mills equations of motion in the presence of colliding shockwave sources.

The ``classical" computation of Eq.~\eqref{spectrum} was performed\footnote{These references took (seemingly) different classical approaches to computing the energy spectrum. While the method in \cite{Spirin:2015wwa} is the same in as our presentation here, \cite{Gruzinov:2014moa} studied the spectrum using the complex Bondi-Sachs news function $C$ within the Fraunhofer approximation of classical radiation theory. The energy flux density at future null infinity $\mathcal{I}^+$ is given in terms of $C$ by $$\frac{d E^{G W}}{d u}=\frac{1}{2 \pi} \int d^2 \Omega\left|\partial_u C\right|^2$$ where $u$ is the retarded time; at null infinity $\mathcal{I}^+$ this becomes $u=t-r$, and the news function $C$ is given by the asymptotic form of Riemann tensor in the post collision spacetime. Fourier transforming this equation w.r.t. $u$ gives $\frac{dE^{\rm GW}}{d\omega}$.} in \cite{Gruzinov:2014moa, Spirin:2015wwa}  and the main features of the spectrum were found to agree in a computation based purely on the amplitudes approach in \cite{Ciafaloni:2015vsa, Ciafaloni:2015xsr}. As we explain below, the result of our paper clearly illuminates why the two approaches {\it should agree}.

The problem of gravitational radiation has an old history and has been studied using several approaches in a variety of scenarios. Early works on computing energy spectrum of gravitational radiation were done in setups which had as ingredients 
\begin{itemize}
    \item nonrelativistic and relativistic scattering with finite boost factor $\gamma$,
    \item  the masses of the colliding particles were nonvanishing,
    \item the impact parameter is large.
\end{itemize}
In \cite{Matzner:1974rd}, Matzner and Nutku used the Weizs\"{ac}ker-Williams method of virtual quanta to compute the spectrum of gravitational radiation emitted from a particle scattering off a Schwarzschild black hole. More definitive investigations of gravitational wave radiation in collision of two relativistic sources dates back to the works of D'Eath \cite{DEath:1976bbo} and by Kovacs and Thorne  \cite{Kovacs:1977uw,Kovacs:1978eu}. These papers computed the frequency and angular distribution of gravitational energy radiated in the regime of small deflection angles (of the trajectories of incoming particles) and small emission angles (of the emitted radiation as measured from the collision axis in the center of mass frame). 

The case of scattering of strictly massless particles was studied more recently in \cite{Gruzinov:2014moa, Spirin:2015wwa, Ciafaloni:2015vsa, Ciafaloni:2015xsr}. It was found that at low enough frequencies ($\omega \lesssim b^{-1}$), the energy spectrum is approximately flat and reaches a constant value at zero frequency. This is famously known as the zero frequency limit (ZFL) \cite{Smarr:1977fy} and is dictated by Weinberg soft graviton theorem \cite{Weinberg:1965nx}. Roughly speaking, the ZFL is obtained from the Weinberg current $$J^{\m\n}_{W}=\kappa \sum_i \eta_i \frac{p_i^\mu p_i^\nu}{p_i \cdot q}$$ (where $p_i$ are (hard) momenta of external lines from which soft graviton of momenta $q$ is emitted) in the construction of the amplitude $\mathcal{M}^{(\lambda)}_W = J^{\m\n}_{W} \varepsilon_{\m\n}$ in Eq.~\eqref{spectrum}. One finds that at $\omega=0$, the angular integrated result for small deflection angles $\theta^2 \sim -4t/s$ takes the form \cite{Weinberg:1972kfs,Gruzinov:2014moa}
\be
\frac{dE^{\rm GW}}{d\omega} = -\frac{4}{\pi}Gt \log\(\frac{s}{-t}\)~.
\ee
This is the small frequency part of the spectrum which is universal in the sense that the details of the internal structure of the scattering objects and that of the scattering process are not important. This regime reliably describes scattering processes at both large and small impact parameters. 

On the other hand, to go away from the ZFL, the details of scattering become important. For Regge scattering at large impact parameters, the energy spectrum is likewise computed by replacing in the above argument the Weinberg current with the Lipatov current $\Gamma^{\m\n}$ defined in Eq.~\eqref{GLV}. The resulting expression was analyzed in detail \cite{Ciafaloni:2015vsa, Ciafaloni:2015xsr} and the results were found to be consistent with the classical computation of the spectrum in \cite{Spirin:2015wwa}. In light of our result, this is to be expected since the radiation amplitude obtained from the classical approach is nothing but Lipatov's emission vertex -- a point that was not emphasized in \cite{Spirin:2015wwa}.

\begin{figure}[ht]
\centering
\includegraphics[scale=1]{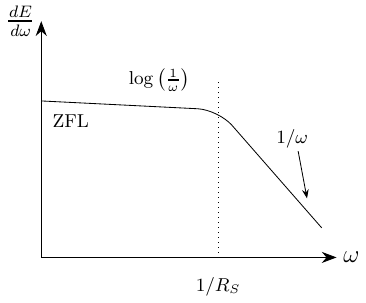}
    \caption{Schematic plot of the frequency spectrum of gravitational radiation in high energy gravitational shockwave collisions at small emission angles \cite{Ciafaloni:2015vsa, Ciafaloni:2015xsr}. The key features of the plot are (1) the zero frequency limit (ZFL) that is governed by Weinberg soft graviton current (2) the nearly-flat logarithmic behaviour in the region $b^{-1}<\omega<R_S^{-1}$ and (3) $1/\omega$ behaviour beyond the $\omega\sim R_S^{-1}$ region that gets modified around $O(b^2/R_S^3)$.}
    \label{spectrumGR}
\end{figure}

In the Regge scattering regime, a unanimous conclusion of these (classical and quantum amplitudes based) studies is the emergence of a  characteristic frequency scale $\omega \sim 1/R_S$ beyond which the spectrum ceases to be flat and assumes a $1/\omega$ behaviour. This is interesting  since even though at large impact parameters one is away from the black hole formation region, the spectrum already shows signs of the characteristic gravitational radius. However this feature of the spectrum leads to a logarithmic UV divergence for the total energy radiated. One can argue that such a divergence can be naturally regulated by the nonlinearities of general relativity (that were largely ignored in these studies) which will smoothly cut off the spectrum at some characteristic frequency. Indeed as mentioned in \cite{Gruzinov:2014moa}, the method used in their work has the shortcoming of neglecting the strong curvature regime thereby excluding a rigorous treatment of the high frequency domain. Under the approximation used in those studies (inelastic scattering at large impact parameters), the $1/\omega$ behaviour is valid until  $\omega \sim b^2/R_S^3$, beyond which a hard cutoff leads to the $1/\omega^2$ behaviour of the spectrum \cite{Gruzinov:2014moa}. (Indications of this profile were already seen in the extrapolations of the finite $\gamma$ results in \cite{Kovacs:1978eu}.) We  depict this schematically in Fig.~\ref{spectrumGR}. 

\begin{figure}[ht]
\centering
\includegraphics[scale=1.0]{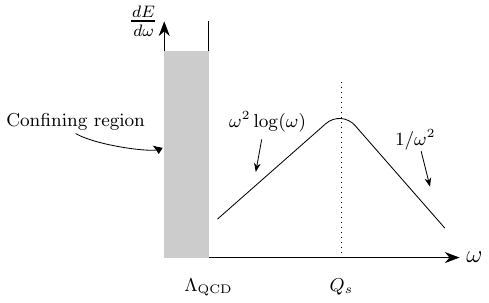}
    \caption{Schematic plot of the classical energy spectrum of gluons emitted in gluon shockwave collisions. The key features are (1) at large $\omega$ the spectrum is dominated by the Weizs\"{a}cker-Williams $1/\omega^2$ behaviour (2) logarithmic behaviour below the $\omega \sim Q_s$ saturation scale which continues till $\omega \sim \Lambda_{\rm QCD}$.}
    \label{spectrumQCD}
\end{figure}

One can contrast these results with the spectrum of gluons emitted in  QCD shockwave collisions schematically shown in Fig. \ref{spectrumQCD}. At high $|\bsk| \sim \omega$ (of the radiated gluon), gluons  have a perturbative tail due to asymptotic freedom (as opposed to gravity which becomes strong in the UV $\sim M_{\rm pl}$) and their energy spectrum effectively follows the corresponding Weizs\"{a}cker-Williams $1/\omega^2$ distribution. This behavior gets modified at the saturation scale $|\bsk|=Q_s$ below which the behaviour is logarithmic in $\omega$ until $\omega\sim \Lambda_{QCD}$ where confinement takes over (unlike gravity which become free in the IR). In both QCD and gravity, one sees the emergence of a saturation scale in the radiation spectrum~\cite{McLerran:1993ka,Dvali:2010jz,Dvali:2021ooc}. 

It would be of great interest to revisit this issue in the dilute-dense approximation (discussed further below) which can potentially allow us to rigorously analyze the spectrum in the high frequency domain. This approximation would include multiple scattering effects of the emitted graviton off the heavy source shown schematically in Fig. \ref{dilute-dense-grav}. The result of the dilute-dense approximation (which includes resummation of  coherent rescattering encoded in the gravitational Wilson line) can lead to  improved UV behavior of the frequency spectrum via an exponentiation of the leading order process; this is precisely what happens in QCD. 

The exponentiation in QCD was already shown in the dilute-dense solution in Eq.~\eqref{dilutedenseQCD}, which  was used to compute the spectrum of emitted gluons. We point out that this computation in gravity is for large impact parameters but taking into account the short distance nonlinear effects of GR which a highly energetic  graviton emitted off the lighter shock will be sensitive to. Finding a detailed form of the high frequency spectrum would be relevant for future gravitational wave observations which include gravitational wave bursts \cite{Garcia-Bellido:2017knh, Garcia-Bellido:2017qal, Morras:2021atg} emitted in close hyperbolic encounters of highly energetic sources like primordial black holes \cite{Escriva:2022duf} or as stochastic backgrounds thereof \cite{Garcia-Bellido:2021jlq,Mukherjee:2021ags}.

While the aforementioned computation is likely within reach, the interesting problem of the gravitational wave spectrum produced near the black hole formation threshold is arguably much more challenging. This threshold is expected to occur at a critical impact parameter $b\sim R_S$ \cite{Thorne:1972ji} and is characterized by deflection angles of order one. This problem in trans-Planckian scattering has seen several attempts in the recent past \cite{East:2012mb, Rezzolla:2012nr, Giddings:2004xy, Yoshino:2005hi, Yoshino:2002tx, Eardley:2002re, Ciafaloni:2017ort} but a satisfactory picture is still elusive. One naively expects that the spectrum of GW radiation near the threshold to be very different due to large numbers of fast moving gravitons in the $2\to N$ final state all of which are multiple scattering with both the shock waves. This is the full dense-dense regime where the efficiency of gravitational radiation (defined as the ratio $E^{\rm GW}/E$ of energy radiated in gravitational waves to the total energy of the process) is a relevant quantity of interest that has been much debated topic in the literature \cite{Pretorius:2007jn,Sperhake:2012me,East:2012mb,Page:2022bem}. It is clearly interesting therefore to obtain a quantitative understanding of the spectrum of gravitational radiation as a function of the impact parameter especially for values where deflection angles are not small and tidal effects become important. To the best of our knowledge this problem has not been studied in detail for the trans-Planckian scattering regime. This is where the dilute-dense generalization of the computation presented in this paper would be crucially relevant that we will outline next. 

\subsection{Dilute-dense generalization and renormalization group evolution}
\label{sec:dilutedense}

\begin{figure}[ht]
\centering
\includegraphics[scale=1.0]{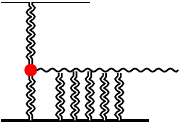}
\caption{The dilute-dense generalization depicting coherent multiple scattering  off the dense source that a highly energetic graviton emitted from the light source will be sensitive to. The double graviton lines represent reggeized graviton propagators that are important for smaller impact parameters.}
\label{dilute-dense-grav}
\end{figure}

We will now comment on the generalization of our result of the Lipatov's emission vertex to the dilute-dense approximation where the high occupancy source $\rho_H$ is resummed to all orders representing its coherent scattering of the radiated graviton field  as illustrated in Fig.~\ref{dilute-dense-grav}. 
An important reason for pursuing this generalization (other than those mentioned in the previous section) is to investigate the gravitational analog of the Balitsky-Kovchegov equation in QCD, as discussed previously in Section~\ref{subsec:Lipatov-gravity}. In the latter case, the nontrivial fixed point of RG evolution corresponds to gluon saturation with an emergent saturation scale $Q_s$; the corresponding RG equation in gravity would describe an analogous saturation of semi-hard gravitons at the emergent scale $b=R_S$. 

Substantial insights towards this goal can be gained from the QCD case which, following \cite{Blaizot:2004wu,Gelis:2005pt}, we briefly describe here. There are essentially three steps involved. The first step is to calculate the evolution of the current of the light shockwave as it passes through the dense shockwave on the other light cone. In the strict eikonal approximation (where the trajectories of the shockwaves do not bend), this evolution is straightforwardly determined by covariant current conservation and results in dressing up the initial current by a light-like Wilson line\footnote{
In the dilute-dense approximation in QCD, the dominant contribution appears to come from multiple $t$-channel gluon exchange between color sources at higher rapidities (with occupancy  $O(1/\alpha_s)$) and  the outgoing gluon. Since the shockwave is at large occupancy, classical effects dominate over the quantum corrections; resummation over coherent multiple gluon exchanges is represented by a Wilson line, which is a solution to the classical equations of motion. On the surface, this power counting dominates over contributions from real emissions and virtual loops. However though these are formally suppressed by $\alpha_s$, they contain large $\alpha_s\log(s/|t|)$ contributions of $O(1)$ in Regge asymptotics and therefore of equal importance as multiple scattering contributions.}.

\begin{figure}[ht]
\centering
\raisebox{-12pt}{\includegraphics[scale=1.0]{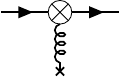}}
=
\raisebox{-0pt}{\includegraphics[scale=1.0]{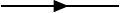}}
+ $\sum_{n=1}^\infty$
\raisebox{-12pt}{\includegraphics[scale=1.0]{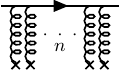}}
\caption{Dressed propagator of a Dirac fermion in the gluon shockwave background represented by the cross. The first term on the r.h.s is the free Dirac propagator. The second term  resums  multiple scatterings of the fermion line on the background, corresponding to the effective vertex in Eq.~\eqref{qcd-fermion-eff-vertex}.}
\label{shockwavepropagator}
\end{figure}

The next step is to linearize the Yang-Mills equations around the dense shockwave background. The equations are inhomogenous as they are sourced by the evolved current determined from the previous step. The last step is to solve these linearized inhomogenous differential equations which requires the knowledge of Green's functions of the wave operator in shockwave background. These propagators were computed in \cite{McLerran:1994vd,
Ayala:1995kg,Balitsky:2001mr}. One needs to solve an eigenvalue differential equation of the form 
\be
\square_{\rm shock} \phi = \lambda \phi \,,
\ee
where $\square_{\rm shock}$ is the Laplacian in the shockwave background; the shockwave Green's function is  then given by 
\be 
G(x,x') = \int d\lambda \frac{\phi_\lambda(x)\bar{\phi}_\lambda(x')}{\lambda - i\epsilon}\,. 
\ee
For instance, the shockwave propagator of a Dirac fermion (in momentum space) takes the  form \cite{McLerran:1998nk}
\be
i S\left(p, p^{\prime}\right)=(2 \pi)^4 \delta^{(4)}\left(p-p^{\prime}\right) i S_0(p)+i S_0(p) \mathcal{A}_q\left(p, p^{\prime}\right) i S_0\left(p^{\prime}\right)~.
\ee
Here $S_0$ is the free Dirac propagator and $\mathcal{A}_q\left(p, p^{\prime}\right)$ the part that encodes all the multiple interaction with the shockwave background and is given by
\be
\label{qcd-fermion-eff-vertex}
\mathcal{A}_q\left(p, p^{\prime}\right)=2 \pi \delta\left(p_{-}-p_{-}^{\prime}\right) \gamma_{-} \epsilon\left(p_{-}\right) \int d^2 \boldsymbol{z} \mathrm{e}^{-i \boldsymbol{z} \cdot\left(\boldsymbol{p}-\boldsymbol{p}^{\prime}\right)}\left(U\left(p_{-}, \boldsymbol{z}\right)-1\right)\,,
\ee
where $U \left(p_{-}, \boldsymbol{z}\right)=P\mathrm{e}^{i g \epsilon\left(p_{-}\right) \rho(\boldsymbol{z})/\square_\perp}$ is the path ordered Wilson line operator (in the same representation of the gauge group as that of the Dirac fermion) and $\epsilon(x)=\theta(x)-\theta(-x)$. The dressed propagator can be pictorially represented as shown in Fig.~\ref{shockwavepropagator}. 
A similar expression exists for the gluon propagator in the shockwave background~\cite{Ayala:1995hx,Ayala:1995kg,Balitsky:2001mr}. 

One anticipates the same to hold for  propagators in the background of gravitational shockwaves. For real scalar fields the gravitational shockwave propagator takes the form
\begin{equation}
i S\left(p_1, p_2\right)=(2 \pi)^4 \delta^{(4)}\left(p_1-p_2\right) i S_0(p_1)+i S_0(p_1) A_{\text {shock}}\left(p_1, p_2\right) i S_0\left(p_2\right)~,
\end{equation}
where $S_0$ is the free scalar propagator and $S_0 AS_0$ is the effective vertex with $A(p_1,p_2)$ given by \cite{deGioia:2022fcn}
\begin{equation}
A_{\text {shock }}\left(p_1, p_2\right)=4 \pi p_2^{-} \delta\left(p_2^{-}-p_1^{-}\right) \int d^2 \bsx e^{i\left(\bsp_2 - \bsp_1\right) \cdot \bsx} \left(e^{-i \frac{\bar{g}\left(\bsx\right)}{2} p_1^{-}}-1\right)~.
\end{equation} 
The exponent appearing in the parenthesis is the gravitational Wilson that we discussed  in Section \ref{sec:3b}. 

The knowledge of the Green's function in the shockwave background is imperative for deriving the BK equation in QCD. The basic idea is that one considers the interaction of a quark-antiquark dipole with the shockwave and consider all dressed one-loop diagrams that result from emission and absorption within the dipole. One such one-loop diagram involves a gluon propagator that undergoes coherent scattering with the shockwave. The fundamental quantity of interest is the two-point correlator of Wilson lines (denoted by $\boldsymbol{S}$) whose one loop correction can be explicitly computed given the gluon shockwave propagator. It turns out that the one-loop correction to this transition amplitude is larger than the tree-level term by a factor of $\alpha_s Y$ where $Y\equiv\log(p^+/P^+)$ is the large phase space factor that comes from the one-loop integral where $p^+$ is the corresponding gluon light cone momentum and $P^+$ is the momentum of the nucleus. For $\alpha_s Y\sim 1$, one needs to resum such contributions to all-loop orders, amounting to exponentiation the one-loop term. From this result, the BK equation (which is an evolution of $\boldsymbol{S}$ in $Y$) can be extracted - see \cite{Gelis:2007kn} for further details.

We expect that these techniques will be applicable to the dilute-dense generalization in gravity. In this paper, we  computed the evolution of EM tensor of a boosted point-like source as it scatters off the dense shockwave. From this result, one can conclude that at large impact parameters the scattering is essentially eikonal and the EM tensor of the light shockwave is dressed by the gravitational Wilson line operator in Eq.~\eqref{VWilson}. However it is evident from Eq.~\eqref{Tijsol1} that this no longer true as one lowers $b$ towards $R_S$ where the EM tensor develops nonvanishing components along the transverse directions. Such terms were already important to consider in the dilute-dilute computation; we therefore expect these to also be relevant for the dilute-dense generalization. This generalization will also incorporate reggeization of semi-hard gravitons encoded in the ${\rho}_{A,B}(\bsq)/\bsq^2$ factors. As discussed at length in Section \ref{subsec:Lipatov-gravity} reggeization occurs due to multi-loop graviton exchanges at order $O(-t/s)$. In order to recover this effect in the shockwave picture, one needs to compute the graviton propagator in the shockwave background and demonstrate that ${\rho}_{A,B}(\bsq)/\bsq^2$ satisfy the graviton Regge trajectory given in Eq.~\eqref{gravitonReggeTrajectory}. The analog in QCD is the derivation of the BFKL equation via the shockwave framework. In gravity we expect this computation to give the evolution equation computed by Lipatov in Eq.~(80) of \cite{Lipatov:1982vv}. A further extension, as previously discussed in Section \ref{subsec:Lipatov-gravity}, would be to compute the gravitational analog of the BK equation and determine its regime of validity relative to the counterpart of the BFKL equation. The computation of the graviton shockwave propagators  will be addressed in a forthcoming publication.

\subsection{Double copy and asymptotic symmetries}

A topic of great recent interest are the constraints imposed on amplitudes in gauge theories and gravity by asymptotic symmetries, the study of which goes by the name celestial holography~\cite{Pasterski:2021raf}. These asymptotic symmetries can be understood as giving rise to soft theorems in gauge theories and gravity and a memory effect that in principle can be measured~\cite{Strominger:2017zoo}. Of interest to us is the concrete analogy between the color memory effect in QCD and the gravitational memory effect~\cite{Pate:2017vwa}. In \cite{Ball:2018prg}, it was shown that the CGC shockwave encodes this color memory effect and that the 2-D dynamics of wee gluons in the transverse plane could, through a stereographic projection, be mapped on to the celestial sphere at null infinity. 

This raises the question whether the dynamics of wee gluons can be understood as Goldstone dynamics on the celestial sphere, as for instance argued for soft photons in \cite{Arkani-Hamed:2020gyp} and for soft gravitons in \cite{He:2023qha}. The broken asymptotic symmetry in this case is that of large gauge transformations spontaneously broken by the shockwave, which as we have seen in Section~\ref{sec:gr-shockwaves} is common to both our QCD and gravity shockwave formulations. In the gravity case, this is the Bondi-van der Burg-Metzner-Sachs (BMS) asymptotic symmetry of supertranslations and superrotations~\cite{Bondi:1962px,Sachs:1962wk}. In this mapping there is the key assumption that the shockwave is static in light cone time $x^+$- the transverse dynamics  hence can be ported to null infinity. While this is manifestly not the case due to confinement, it can be argued that for perturbative processes on short time scales the  assumption may be valid. However a more subtle issue is the semi-classical nature of the condensate, where the finite energy levels of $O(1/N)$ (note $N= 1/\alpha_s$ is the occupancy of modes) imply the shockwave decays on a time scale characterized by the Goldstone decay constant. 

In the Regge limit of QCD, the symmetries satisfied by the emergent semi-classical color charge densities (represented for instance by $\rho_{H}$ in our notation) are those of the largest symmetric sub-group of $SU(3)$-see for instance the discussion in \cite{Jeon:2004rk}. The corresponding Goldstone modes are the reggeized gluon fields we discussed that have frequencies $\omega \sim Q_S$. Since this is not $\omega\rightarrow 0$, as would be the case if they were truly asymptotic, it would be interesting to explore whether the geometric interpretation in terms of BMS symmetries is robust. This has been argued to be the case in the Black Hole Quantum Portrait framework~\cite{Dvali:2015rea,Averin:2016hhm}.
Indeed in \cite{Dvali:2021ooc}, it was shown that the parametric dependence of the reggeized graviton-reggeized graviton-graviton vertex and that of its QCD counterpart were identical; it is important to note that this correspondence is only applicable when $\alpha N = O(1)$. Further studies exploring the impact of asymptotic symmetries on the classical double copy are needed.

\section{Summary}
\label{sec:summary}

In this work, we explored a quantitative connection between $2\rightarrow N$ scattering in QCD and Einstein gravity in Regge asymptotics. Despite the clear differences between the two theories, the dynamics of multiparticle production has a number of common features in the Regge limit. Specifically, the dynamics is controlled by the copious production of wee partons in both theories. In QCD, this leads to the well known phenomenon of gluon saturation which is quantitatively described by the Color Glass Condensate EFT. In gravity, a parallel development has been to describe black holes as saturated self-bound states of wee gravitons - the Black Hole Quantum N-Portrait (BHNP) framework. The universality between quantum states in the regime where the wee parton occupancy is of the order of the inverse of the coupling in Eq.~\eqref{effectiveCoupling} was noted previously in \cite{Dvali:2021ooc}. 

As we discussed, the CGC description of saturated gluon states is highly developed with the 
evolution of the nonperturbative dynamics of wee partons described by nonlinear RG equations that are known to next-to-leading logs in $x$ accuracy and recovers  the RG equations of perturbative QCD in the UV. The elements of the CGC framework are semi-classical fields that can be mapped on to the language of reggeon field theory (inspired by diagrammatic computations) with a clear correspondence established to its building blocks, reggeization and the Lipatov vertex. Since these are the very same building blocks in perturbative discussions of $2\rightarrow N$ scattering in gravity, it is natural to likewise consider that whether strong field techniques developed in the CGC EFT might be applicable and provide a complementary framework to that developed by ACV which employs the diagrammatic language of reggeon field theory. 

A remarkable additional feature of the two theories that goes beyond the above stated analogies and possible universality is that of the BCJ/classical double copy that has emerged as a powerful tool for quantitative computations. We noted that a particular Regge double copy between the Lipatov vertex in QCD and the effective gravitational central emission vertex was established by Lipatov over 40 years ago. A central result of our paper is that this double copy relation can be recovered in our semi-classical approach in gravitational shock wave collisions at large impact parameters. This derivation parallels the extraction of the QCD Lipatov vertex in the scattering of gluon shockwaves performed previously. Our approach also provides fresh insight from the semi-classical perspective into how and why the QED bremsstrahlung vertex enters into the double copy. The QCD example also suggests a useful classification of shockwave scattering into dilute-dilute, dilute-dense and dense-dense regimes. 

Our results for gravitational shockwave scattering also suggest that one can construct dressed gravitational shockwave propagators with an analogous structure to QCD shockwave propagators. These can then be employed in a semi-classical derivation of graviton reggeization along the lines of gluon reggeization. If successful, this will allow us to construct a nonlinear RG description of multiparticle production as a function of impact parameter and rapidity. The formation of a black hole would then be understood as a nontrivial fixed point of the RG flow. In QCD, the nontrivial fixed point corresponds to the classicalization and unitarization of the $2\rightarrow N$ amplitude; this is also the conjecture in the BHNP and a quantitative derivation in the language of RG flow is desirable. A conceptual issue that would be important to resolve in this regard is how the Goldstone picture of asymptotic symmetries that are broken by both graviton and gluon shockwaves applies at finite frequencies. This is relevant to understanding the emergence of a universal area law in the dynamics of wee partons. 

Not least, we addressed how our semi-classical framework relates to the large body of work on gravitational shockwave scattering. This is of course of great importance in understanding gravitational wave radiation in particular with respect to uncovering possible signatures of RG evolution and black hole formation in the radiation spectrum. We plan to undertake a quantitative study of such signatures in future work.


\begin{acknowledgments}
We thank Gia Dvali and Gabriele Veneziano for valuable discussions that have significantly influenced this work. We thank Simon Caron-Huot and Grisha Korchemsky for useful comments on aspects of the extensions of this work. R.V would also like to thank Agustin Sabio Vera for clarifications of his work on the Lipatov double copy. 

R.V is supported by the U.S. Department of Energy, Office of Science under contract DE-SC0012704 and within the framework of the SURGE Topical Theory Collaboration. He acknowledges partial support from an LDRD at BNL. R.V was also supported at Stony Brook by the Simons Foundation as a co-PI under Award number 994318 (Simons Collaboration on Confinement and QCD Strings). He thanks the DFG Collaborative Research Center SFB 1225 (ISOQUANT) at Heidelberg University for hospitality and support during the course of this work. 

H.R is a Simons Foundation postdoctoral fellow at Stony Brook supported under Award number 994318. H.R acknowledges support by the ERC Starting Grant 853507 and thanks the  EIC Theory Institute at Brookhaven National Lab for its hospitality when this project was nearing its completion. 
\end{acknowledgments}
\appendix


\section{Gluon shockwave scattering in QCD}
\label{appA}

The classical Yang-Mills (YM) equations of motion in the presence of external currents are
\begin{align}
D_\m F^{\m\n}=J^\n~,
\end{align}
where the field strength is defined as $F_{\m\n}=\p_\m A_\n-\p_\n A_\m+ig[A_\m,A_\n]$ and the current $J^\m$ is covariantly conserved: $ D_\m J^\m = 0$. 
The action of the covariant derivative on an adjoint field $F$ is given by
\be
D_\m F = \p_\m F - ig [A_\m, F]~.
\ee
A gluon shockwave is generated by the  current
\begin{equation}
J^{\nu,a}=g \,\delta^{\nu+} \delta\left(x^{-}\right) \rho_H \,T^a\left(\boldsymbol{x}\right)~,
\end{equation}
where $T^a$ is a generator of the color algebra. This form is analogous to the form of the energy-momentum tensor Eq.~\eqref{EMtensor1} in gravity. It is straightforward to verify that the exact solution, $\bar{A}_-^a$ , to the YM equations with this source is given by
\be
\label{gShockBgnd1}
\bar{A}_-^a = -g \, \delta\left(x^{-}\right) \frac{\rho_{H}\left(\boldsymbol{x}\right)}{\square_\perp}\,T^a~.
\ee
This form admits a precise double copy relation to the gravitational shockwave in the coordinate system given in Eq.~\eqref{denseBgnd1} and Eq.~\eqref{backgroundg} through a color-kinematic replacement made precise in \cite{Saotome:2012vy}. In this Appendix, we shall describe the scattering of two such shockwaves in the dilute-dilute approximation and show how the QCD Lipatov vertex arises. 

\subsection*{\label{sec:level2} Gluon shockwave collisions and the Lipatov vertex}
\label{appC1}

The approach for solving this problem is similar to the one in gravity. One begins by linearizing the YM equations over the shockwave background Eq.~\eqref{gShockBgnd} which is  linear in the light cone gauge:
\be
A_+=0~,
\ee
In this gauge, the YM equations can be written as 
\begin{align}
\label{LinYM1}
\begin{split}
    &-\p_+ \p_\m A^\m+ ig\[A_\m, \p_+ A^\m\] =J_+~,\\
    &\square A_- -\p_- \p_\m A^\m - ig \[\p_\m A^\m,A_-\] - 2ig \[A^\m,\p_\m A_-\] + ig\[A_\m, \p_- A^\m\] = J_-~,\\
    &\square A_i -\p_i \p_\m A^\m - ig \[\p_\m A^\m,A_i\] - 2ig \[A^\m,\p_\m A_i\] + ig\[A_\m, \p_i A^\m\] = J_i~.
\end{split}
\end{align}
A straightforward linearization $A_\m=\bar{A}_\m+a_\m$ of these equations gives the set of equations
\begin{align}
\label{LinYM2}
\begin{split}
    &-\p_+ \p_\m a^\m=J_+\\
    &\square a_- -\p_- \p_\m a^\m - ig \[\p_\m a^\m,A_-\] - 2ig [\bar{A}_-,\p_+a_-] + 2ig \[a_i,\p_i \bar{A}_-\] = J_-~,\\
    &\square a_i -\p_i \p_\m a^\m - 2ig \[\bar{A}_-,\p_+ a_i\] = J_i~.
\end{split}
\end{align}
The first equation is a constraint relation which relates $a_i$ to $a_-$. This implies that the equations for $a_-$ and $a_i$ are not independent. The consistency of the set of equations is represented by 
\begin{align}
\label{gaugeCur}
\p_\m J^\m = i g \[\bar{A}_-, J_+\] ~.
\end{align}
which is nothing but covariant current conservation. This discussion is entirely analogous to that in gravity. 

Next, we incorporate the effect of the other incoming shockwave. This is generated by the current
\be
J_{+,a} = g \delta(x^+)\rho_L\(\bsx\)T_a\,.
\ee
Plugging this relation into Eq.~\eqref{LinYM2} and first solving to linear order in $\rho_L$ but zeroth order in $\rho_H$ gives us the gauge field for the second shock in light cone gauge:
\be
\label{gShockBgnd2}
a_{i,a} = g\,\Theta(x^+)\frac{\p_i \rho_L}{\square_\perp}T_{a}+O(\rho_L\,\rho_H)~,\qquad a_-= O(\rho_L\,\rho_H)~.
\ee

We now need to determine the form of the current post-collision. In QCD, the problem of finding the evolved current is considerably simpler than that in gravity. The reason is that in QCD one can solve for the currents using simply the conservation law in the strict eikonal approximation where the initial current $J_+$ of the second shock does not develop any components along the transverse and $-$ longitudinal direction. Then in the dilute-dilute approximation, Eq.~\eqref{gaugeCur} can be solved for the correction to $J_+$ by inserting the leading order results on the r.h.s. Using the Lie algebra relation $[T_a,T_b]=if_{abc} T_c$, we get
\begin{align}
\label{currEv}
J_{+,c} & = g^3 \delta(x^+)\Theta(x^-)\frac{\rho_{H}}{\square_\perp}\rho_{L}T^a T^b f_{abc} ~,
\end{align}

Plugging the results from Eqs.~\eqref{gShockBgnd2} and \eqref{currEv} into the equations of motion of $a_i$ and $a_-$ in Eq.~\eqref{LinYM2} gives us (at order $O(\rho_L\rho_H)$)
\begin{align}
    &\square a_{i,c} = -g^3\(\Theta(x^+)\Theta(x^-) \p_i\(\frac{\rho_H}{\square_\perp}\rho_L\)-2\delta(x^+)\delta(x^-) \frac{\rho_H}{\square_\perp} \frac{\p_i\rho_L}{\square_\perp}\)T^a T^b f_{abc}~,\\[5pt]
    &\square a_{-,c}= 2g^3\Theta(x^+)\delta(x^-) \(\frac{\p_i\rho_H}{\square_\perp} \frac{\p_i\rho_L}{\square_\perp}\)T^a T^b f_{abc}
\end{align}

Upon taking the Fourier transforms of these equations, and putting the momenta of the emitted gluon on-shell, we find
\begin{align}
\label{gaugesol}
    a_{i,c}(k) &= -\frac{2ig^3}{k^2+i\epsilon k^-}\int \frac{d^2\bsq_2}{(2\pi)^2}\(q_{2i}-k_i\frac{\bsq_2^2}{\bsk^2}\)\frac{\rho_H}{\bsq_1^2}\frac{\rho_L}{\bsq_2^2}T^a T^b f_{abc}~,\\[10pt]
    a_{-,c}(k) &= -\frac{2ig^3}{k^2+i\epsilon k^-}\int \frac{d^2\bsq_2}{(2\pi)^2} \frac{(\bsq_1\cdot \bsq_2)}{k_+}\frac{\rho_A}{\bsq_1^2}\frac{\rho_B}{\bsq_2^2}T^a T^b f_{abc}~.
\end{align}

From these expressions, we can now extract the QCD Lipatov vertices in the light cone gauge:
\begin{align}
    C_i=-2\(q_{2i}-k_i\frac{\bsq_2^2}{\bsk^2}\)~,\qquad C_- = -2 \frac{(\bsq_1\cdot \bsq_2)}{k_+}~.
\end{align}
The above result satisfies $\p^\m C_\m=0$. 
This computation can be straightforwardly generalized to all orders in $\rho_H$ - as performed in \cite{Blaizot:2004wu, Gelis:2005pt}.


\section{GR review and linearization of Einstein's equations}
\label{appB}


\subsection{GR review and conventions}
\label{appA1}
The action for Einstein-Hilbert gravity is 
\be
\label{EHAction}
S_{\text {grav }}=-\frac{1}{2 \kappa^2} \int d^{4} x \sqrt{-g} R\,,
\ee
where $g=\operatorname{det}\left(g_{\mu \nu}\right)$ is the determinant of the metric tensor and $\kappa$ is proportional to the Newton constant $G$
\be
\kappa^2=8 \pi G = \frac{8\pi}{M_p^2}\,.
\ee
It has mass dimension $[\kappa]=-1$ in natural units ($\hbar = 1$). The scalar curvature $R$ is related to the Ricci curvature tensor $R_{\m\n}$ by the contraction of its indices $R=R_{\mu \nu} g^{\mu \nu}$. The Ricci tensor is in turn given by the contraction of components of the Riemann tensor:
\be
R_{\mu \nu}=R_{~\mu\sigma \nu}^{\sigma}\,.
\ee
We are working in conventions where the Riemann curvature tensor is given by the commutator (following the same convention as Wald's book but opposite to Weinberg's convention, where there is a minus sign on the r.h.s)
\be
\label{ReimannDef}
[\nabla_\sigma,\nabla_\rho]V_\nu =  V_\m R^\m{}_{\n\rho\sigma}~,
\ee
with $V$ being an arbitrary 4-vector and $\nabla$ the covariant derivative whose action on $V$ is defined as 
\be
\nabla_{\mu} V_{\nu}= V_{\n;\m}=\partial_{\mu} V_{\nu}-\Gamma_{\mu \nu}^{\rho} V_{\rho}~,\qquad \Gamma_{\mu \nu}^{\rho}=\frac{1}{2} g^{\rho \sigma}\left(\partial_{\mu} g_{\sigma \nu}+\partial_{\nu} g_{\mu \sigma}-\partial_{\sigma} g_{\mu \nu}\right)~.
\ee
Eq. \eqref{ReimannDef} then gives the Riemann tensor in terms of the following combination of Christoffel symbols $\Gamma$ and its derivatives:
\be
R_{~\mu \alpha \beta}^{\sigma}=\partial_{\alpha} \Gamma_{\mu \beta}^{\sigma}-\partial_{\beta} \Gamma_{\mu \alpha}^{\sigma}+\Gamma_{\mu \beta}^{\rho} \Gamma_{\rho \alpha}^{\sigma}-\Gamma_{\mu \alpha}^{\rho} \Gamma_{\rho \beta}^{\sigma}~.
\ee
Varying the Einstein-Hilbert action with respect to the metric gives the Einstein field equations
\be
\label{EinsteinEq}
R_{\mu \nu}-\frac{1}{2} g_{\mu \nu} R = \kappa^2 T_{\m\n}~,
\ee
where the energy-momentum tensor $T_{\m\n}$ is included in the r.h.s to account for possible matter contributions. Denoting the matter action by $S_m$, one {\it defines} $T_{\m\n}$ by the metric variation of $S_m$:
\be
\delta S_{m}=\frac{1}{2} \int d^{4} x \sqrt{-g} ~\delta g_{\mu \nu} T^{\mu \nu}~.
\ee
The Bianchi identity of the Riemann tensor $R_{\lambda \mu \n \kappa ; \eta}+R_{\lambda \mu \eta \n ; \kappa}+R_{\lambda \mu \kappa \eta ; \n}=0$ implies that the energy-momentum tensor is covariantly conserved,
\be
\nabla_{\mu} T^{\mu}_{~\nu}=0~.
\ee
The action for a massless point particle is
\be
\label{ppaction}
S_m = \frac12 \int d\l ~\eta^{-1} \dot{X}^\m\dot{X}^\n g_{\m\n}(X)~,
\ee
where dots denote differentiation w.r.t. $\l$, $\eta$ is the determinant of the metric along the worldline that can be gauged away and $g_{\m\n}(X)$ is the metric tensor of the background that the worldline $X(\s)$ is probing. Functionally differentiating Eq.~\eqref{ppaction} with respect to the spacetime metric $g_{\m\n}$ gives the energy-momentum tensor in Eq.~\eqref{PPEMT}.


\subsection{Linearization of Einstein's equations}
\label{appA2}
Here we present the linearization of Einstein's equations around the shockwave background in Eq.~\eqref{denseBgnd1}. One begins by expanding out the metric as in Eq.~\eqref{perturb}. We will be working throughout in the light cone gauge 
\be
h_{\mu+}=0~.
\ee
After straightforward (though lengthy) algebra we find that the linearization of Ricci scalar is given by ($h\equiv h_{ij}\delta_{ij}$ below)
\begin{align}\label{deltaR}
\delta R &=-\bar{g}_{--}\p_+^2 h+\p_+^2 h_{--} - 2\p_+\p_i h_{-i} +2\p_+\p_- h + \(\p_k\p_l h_{kl}-\square_\perp h \)~,
\end{align}
and the  the linearization of Einstein's field equation gives (we have suppressed factors of $\kappa^2$ for clarity)
\begin{subequations}
\label{EinsteinsEq}
\begin{align}
\label{EinsteinsEq++}
&\frac12 \p_+^2 h = T_{++}~,\\
\label{EinsteinsEq+-}
&\frac12\(\p_+^2 h_{--}-\p_+\p_j h_{-j} +\p_+\p_-h \)  -\frac12\delta R = T_{+-}~,\\
\label{EinsteinsEq+i}
&\frac12\(\p_+^2 h_{-i} - \p_+ \p_j h_{ij} + \p_+ \p_i h \) = T_{+i}~,\\
\label{EinsteinsEqij}
&\frac12 \bigg(\bar{g}_{--}\p_+^2 h_{ij}-2\p_+\p_-h_{ij}+\p_+\p_ih_{-j} + \p_+\p_jh_{-i}  + \square_\perp h_{ij}-\p_k\p_ih_{kj}-\p_k\p_jh_{ki}+\p_i\p_j h  \bigg)+\frac12 \delta_{ij} \delta R = T_{ij}~,\\
\label{EinsteinsEq--}
&\frac14 \bigg(2\bar{g}_{--} \p_+^2 h_{--} -\p_-(\bar{g}_{--})\p_+h+2\square_\perp h_{--} -4\p_-\p_jh_{-j}+2\p_-^2 h\no\\
&+2\(\p_i\p_j \bar{g}_{--}\)h_{ij}  + 2 \(\p_i\bar{g}_{--}\) \p_jh_{ij}  +2\(\p_i\bar{g}_{--}\) \p_+h_{-i}  -  \p_i\(\bar{g}_{--}\)\p_ih \bigg)-\frac12 \bar{g}_{--} \delta R = T_{--}~,\\
\label{EinsteinsEq-i}
&\frac12\bigg(\bar{g}_{--}\p_+^2 h_{-i} +\square _\perp h_{-i} - \p_i \p_jh_{-j} + \p_i\p_- h - \p_j\p_-h_{ij} - \p_+\p_-h_{-i}\no\\
&+\p_+\p_ih_{--}  -\frac12 \p_+h ~\p_i \bar{g}_{--}  +\p_+h_{ij} \p_j \bar{g}_{--}\bigg) = T_{-i}~.
\end{align}
\end{subequations}
These are respectively the $++,+-,+i, ij, --,-i$ components of Einstein's equations. Furthermore, linearizing the covariant energy-momentum conservation equations over the background in Eq.~\eqref{backgroundg} give us
\begin{subequations}
\label{EMConservation}
\begin{align}
\label{EMConservation1}
 &\p_-T_{++}+\p_+T_{+-}-\p_i T_{i+}={g}_{--} \p_+T_{++} ~,\\
 \label{EMConservation2}
 &\p_-T_{+-} + \p_+T_{--}-\p_iT_{i-}= {g}_{--}\p_+T_{+-}+\frac12 \(\p_-{g}_{--}\)T_{++} + \frac12 \square_\perp {g}_{--} \frac{1}{\p_+}T_{++}~,\\
 \label{EMConservation3}
&\p_+T_{-i}+\p_-T_{+i}-\p_j T_{ji}= \frac12 \(\p_i {g}_{--}\)T_{++}+{g}_{--}\p_+T_{+i}~.
\end{align}
\end{subequations}
The conservation equations \eqref{EMConservation} are not independent of Einstein's equations \eqref{EinsteinsEq}. The latter implies the former (which, as mentioned earlier, is a consequence of the Bianchi identity). This can be seen upon diagonalizing \eqref{EinsteinsEq} which we briefly outline below. 

From the $++$ component we simply get
\be
\label{traceEq}
h=\frac{2}{\p_+^2} T_{++}~.
\ee
The $+i$ equation Eq.~\eqref{EinsteinsEq+i} together with Eq.~\eqref{traceEq} gives a constrain relation 
\begin{align}
\label{constr1}
\p_j h_{ij}&=\p_+h_{-i}+\frac{2\p_i}{\p_+^2} T_{++}-\frac{2}{\p_+}T_{+i}
\end{align}
Next, a simple manipulation reveals that the $+-$ equation Eq.~\eqref{EinsteinsEq+-} together with Eq.~\eqref{traceEq} and Eq.~\eqref{constr1} gives the first energy-momentum conservation equation Eq.~\eqref{EMConservation1}. Upon using Eqs.~\eqref{traceEq}, \eqref{constr1} and \eqref{EMConservation1}, the trace of $ij$ component Eq.~\eqref{EinsteinsEqij} is found to give a second constraint relation,
\begin{align}
\label{constr2}
 \p_+^2 h_{--} = \p_i\p_j\tilde h_{ij} -\frac{\square_\perp}{\p_+^2} T_{++}  + T  + 2T_{+-} -\bar{g}_{--} T_{++}
\end{align}
where $\tilde{h}_{ij}$ is defined to be the traceless combination
\be
\tilde h_{ij} \equiv h_{ij}-\frac12 \delta_{ij}h~.
\ee
Inserting Eq.~\eqref{constr2} back into Eq.~\eqref{EinsteinsEqij} finally gives the diagonalized equation for $\tilde{h}_{ij}$: 
\begin{align}
\begin{split}
\label{EinsteinsEqij1}
 \bar{g}_{--}\p_+^2 \tilde h_{ij}-2\p_+\p_-\tilde h_{ij} + \square_\perp \tilde h_{ij}=& \(2\p_i\p_j-\square_\perp \delta_{ij}\)\frac{1}{\p_+^2}T_{++}-2\delta_{ij} \(\bar{g}_{--} T_{++} - \frac{\p_-}{\p_+}T_{++} \)\\
&+2\delta_{ij} T_{+-} +2T_{ij}-\delta_{ij}T-\frac{2}{\p_+} \(\p_iT_{+j}+\p_jT_{+i}\)~.
\end{split}
\end{align}
Likewise, one can get the simplified form of the equations of motion for the remaining components $h_{i-}$ and $h_{--}$ (which couple to $\tilde{h}_{ij}$). We record these below. For the $h_{-i}$ we get
\begin{align}
\label{EinsteinsEq-i1}
\bar{g}_{--}\p_+^2 h_{-i} +\square _\perp h_{-i} -2 \p_+\p_-h_{-i} +\p_+\tilde h_{ij} \p_j \bar{g}_{--} = 2T_{-i}-\frac{2\p_-}{\p_+} T_{+i}-\frac{\p_i}{\p_+}\(\bar{g}_{--}T_{++} -\frac{2\p_-}{\p_+}T_{++} + T \)
\end{align}
whereas for the $h_{--}$, we get
\begin{align}
\begin{split}
\label{EinsteinsEq--1}
&\bar{g}_{--} \p_+^2 h_{--} +\square_\perp h_{--}  -2\p_+\p_- h_{--}  + \(\p_i\p_j \bar{g}_{--}\)\tilde h_{ij}  + 2 \(\p_i\bar{g}_{--}\) \p_j\tilde h_{ij}  \\[5pt]
&=\p_-(\bar{g}_{--})\frac{1}{\p_+}T_{++}-  \(\square_\perp \bar{g}_{--}\)\frac{1}{\p_+^2}T_{++}  + \(\p_i\bar{g}_{--}\)\frac{\p_i}{\p_+^2} T_{++} + \frac{2\p_-}{\p_+}\( \frac{\p_-}{\p_+}T_{++}  -\bar{g}_{--}  T_{++} \)\\[5pt]
&+\(\bar{g}_{--} \)^2 T_{++}+ 2T_{--}+ \bar{g}_{--} \(T - 2T_{+-}\) - \frac{2\p_-}{\p_+} T- \(\p_i\bar{g}_{--}\) \frac{2}{\p_+}T_{+i}~.
\end{split}
\end{align}
We emphasize that the equations for $h_{i-}$ and $h_{--}$ cannot be independent of the equation for $\tilde h_{ij}$ since the two constraint relations Eq.~\eqref{constr1} and Eq.~\eqref{constr2} relates them. The two constraints and the three equations of motion Eqs.~\eqref{EinsteinsEqij1}, \eqref{EinsteinsEq-i1} and \eqref{EinsteinsEq--1} should form a consistent system of equations. One finds that this consistency is met only when energy-momentum conservation (Eq.~\eqref{EMConservation}) is satisfied. Given the complicated form of these equations, it is useful to perform this (tedious) sanity check to ensure that the  equations obtained are free of any errors. 


\section{Evolution of the energy-momentum tensor in shockwave collisions}
\label{appC}
Here we provide details behind the result Eq.~\eqref{EMlower} for the evolved EM tensor in the dilute-dilute approximation. As mentioned in the main text, in the point particle approximation (consistent with the dilute-dilute approximation), thanks to the formula Eq.~\eqref{PPEMT}, the problem of computing the energy-momentum tensor reduces to the problem of computing null geodesics. These can be computed from the geodesic equation Eq.~\eqref{geodesicEq} with the appropriate initial conditions. The solution given in Eq.~\eqref{geodesicSol} is then used to reconstruct the energy-momentum tensor. 

From Eq.~\eqref{PPEMT}, we find the result
\be
\label{Tijsol1}
T^{\m\n} = \m_L \dot{X}^\mu(X^-)\dot{X}^\nu(X^-) \delta(x^+-X^+)\delta^{(2)}\(\bsx-\bsX\)\,,
\ee
where, from \eqref{geodesicEq} we have
\begin{align}
&\dot{X}^- = 1~,\qquad \dot{X}^i = -  \kappa^2\m_H \Theta(X^-) \frac{\p_i\rho_H(\bsb)}{\square_\perp}~,\qquad \dot{X}^+  = -  \kappa^2\m_H \delta(X^-) \frac{\rho_H(\bsb)}{\square_\perp} +\frac{\kappa^4 \m_H^2}{2} \Theta(X^-) \(\frac{\p_i\rho_H(\bsb)}{\square_\perp}\)^2\,.
\end{align}
Eq. \eqref{Tijsol1} is the exact result for the evolved energy-momentum tensor of particle (2) probing the shockwave background Eq.~\eqref{denseBgnd1} of particle (1). However because we are working the dilute-dilute approximation, the relevant expression for us is the expansion of Eq.~\eqref{Tijsol1} to order $\m_H\m_L$. The nonvanishing components of $T^{\m\n}$ are
\begin{align}
\begin{split}
T^{--} &= \m_L\delta(x^+)\delta^{(2)}(\bsx-\bsb) + \kappa^2\m_H\m_L \Theta(x^-)\[ \delta'(x^+) \frac{\rho_H(\bsb)}{\square_\perp} \delta^{(2)}(\bsx-\bsb)+ x^-\delta(x^+) \frac{\p_i\rho_H(\bsb)}{\square_\perp}\p_i \delta^{(2)}(\bsx-\bsb) \]\,,\\[5pt]
T^{-+} &= -  \kappa^2\m_H\m_L \delta(x^+)\delta(x^-) \frac{\rho_H(\bsb)}{\square_\perp} \delta^{(2)}(\bsx-\bsb)~, \qquad T^{-i} = -  \kappa^2\m_H\m_L \delta(x^+)\Theta(x^-) \frac{\p_i\rho_H(\bsb)}{\square_\perp}\delta^{(2)}(\bsx-\bsb)\,.
\end{split}
\end{align}
All the other components are of higher order. This expansion is valid only for large enough impact factors $b$ and small enough times $x^-$. 

At this stage, this result can be generalized to a beam of massless particles moving in the $-z$ direction at various impact factors by simply coarse graining $\delta^{(2)}(\bsx-\bsb)\to \rho_L(\bsx)$. Although this generalization misses various ``shearing" and ``expansion" effects of the null geodesic congruence, for our purposes this replacement suffices. Another limitation of the point-particle approximation is that under generalization to extended transverse sources, it will miss terms of the form $\rho_H \rho_L$. In the point particle approximation such a term is proportional to $\delta^{(2)}(\bsx)\delta^{(2)}(\bsx-\bsb)=\delta^{(2)}(\bsb)\delta^{(2)}(\bsx-\bsb)$ which vanishes for nonvanishing impact parameters. In the generalized setup, such terms can be present as will be seen below. 

In the dilute-dilute approximation, we also need to know the corrections to the energy-momentum tensor of particle (1) due to the gravitational influence of particle (2). In  light cone gauge, the metric that particle (1) sees is the (linearized) metric created by particle (2) (whose leading order energy-momentum tensor is $T^{--}=\m_L\delta(x^+)\rho_L$) which is given by 
\begin{align}
\label{metric2LC}
    ds^2 = 2dx^+dx^- - \delta_{ij} dx^idx^j + 2\kappa^2\m_Lx^+\Theta(x^+)\frac{\p_i\p_j}{\square_\perp}\rho_L ~.
\end{align}
This follows from the previously computed result in Eq.~\eqref{htij1}. To compute null geodesics in this metric, we first compute the connection coefficients. These are found to be
\be
\Gamma^-_{ij} = -\frac12 \p_+\p_i\p_j g_{++}~,~~~~\Gamma^i_{+j} =-\frac12 \p_+\p_i\p_j g_{++}~,~~~~\Gamma^i_{jk} =-\frac12\p_i\p_j\p_k g_{++}~,\qquad g_{++} = 2\kappa^2\m_L \delta(x^+) \frac{\rho_L}{\square_\perp}~.
\ee
The corresponding geodesic equations are
\begin{align}
&\ddot{X}^+=0~,\qquad \ddot{X}^-+\Gamma^-_{ij}\dot{X}^i\dot{X}^j=0~,\qquad \ddot{X}^i+\Gamma^i_{jk}\dot{X}^j\dot{X}^k+2\Gamma^i_{+j}\dot{X}^+\dot{X}^j=0~.
\end{align}
In the dilute-dilute approximation, it is sufficient for us to solve this equation to linear order in $\rho_L$. By inspection, it can be seen that the only solution with the appropriate boundary condition is $X^+=\lambda, X^-=0, \bsX = \bsb $. As such, in  light cone gauge there is no correction to the energy-momentum tensor of particle (1) in the dilute-dilute approximation. Hence the total energy-momentum tensor after the collision (with all components upper) is
\begin{align}
\label{EMupper}
\begin{split}
T^{--} &= \m_L\delta(x^+)\rho_L + \kappa^2\m_H\m_L \Theta(x^-)\[ \delta'(x^+) \frac{\rho_H}{\square_\perp} \rho_L+ x^-\delta(x^+) \frac{\p_i\rho_H}{\square_\perp}\p_i\rho_L \]~,\qquad T^{++} = \sqrt{s_1}\delta(x^-)\rho_H\\[5pt]
T^{-+} &= -  \kappa^2\m_H\m_L \delta(x^+)\delta(x^-) \frac{\rho_H}{\square_\perp} \rho_L~,\qquad\qquad T^{-i} = -  \kappa^2\m_H\m_L \delta(x^+)\Theta(x^-) \frac{\p_i\rho_H}{\square_\perp}\rho_L\,.
\end{split}
\end{align}
Upon lowering the components with respect to the leading order metric  
\begin{align}
\label{metric2LCcorrected}
    ds^2 =~&~ 2dx^+dx^- - \delta_{ij} dx^idx^j +2\kappa^2\sqrt{s_1} \delta(x^-)\frac{\rho_H}{\square_\perp}(dx^-)^2+ 2\kappa^2\m_Lx^+\Theta(x^+)\frac{\p_i\p_j}{\square_\perp}\rho_L dx^idx^j+O(\rho_H\rho_L)~,
\end{align}
we find the result in Eq.~\eqref{EMlower} of the main text. One can check the consistency of this result by verifying energy-momentum conservation (Eq~\eqref{EMConservation}). We find that Eq.~\eqref{EMlower} violates energy-momentum conservation by a term proportional to $\rho_H\rho_L$. For point particle sources this corresponds to $\rho_L \rho_H = \delta(\bsx)\delta(\bsb-\bsx) = \delta(\bsb)\delta(\bsb-\bsx) $ which therefore vanishes for our regime of interest $b\gg R_S$. However, if one adds to $T_{+-}$ in Eq.~\eqref{EMlower} the term 
\be
\label{T+-extra}
\kappa^2\m_H\m_L  \Theta(x^+)\Theta(x^-) \rho_H \rho_L~,
\ee
it is curious to note that the energy-momentum conservation is restored even for vanishing impact parameter (albeit a regime that cannot be probed in the approximation we are working in). Conversely, energy-momentum conservation enforces that $T_{+-}$ contains such a term.


\section{Details of the Lipatov vertex and Fourier transforms}
\label{appD}

Our convention for the Fourier transform is (to avoid clutter, we denote the function and its FT transform by the same symbol)
\be
f(x) = \int \frac{d^dk}{(2\pi)^d} e^{ik_\m x^\m} f(k)~,\qquad f(k) = \int d^dx~ e^{-ik_\m x^\m} f(x)\,.
\ee
This implies $\p_\m \to i k_\m$ and $\Theta(x^\m) \to -i/k_\m$ (which follows from $\p_\m\Theta(x^\m)=\delta(x^\m)$). Finally, the FT of a product of functions $f(x)g(x)$ is
\be
F[f(x)g(x)](k) \equiv \int d^dx~ e^{-ik_\m x^\m} f(x)g(x) = \int \frac{d^dq}{(2\pi)^d}f(k-q)g(q)~.
\ee
Using these conventions we next perform the Fourier transform of equations Eqs.~\eqref{EinsteinsEq-i1}, \eqref{EinsteinsEq--1} in the dilute-dilute approximation. (The analysis for Eq. \eqref{EinsteinsEqij1} was presented in the main text.)

Inserting the leading order result for $\tilde{h}_{ij}$ (from Eq.~\eqref{htij1}) and the energy-momentum tensor, a simple algebra gives (where we restored factors of $\kappa^2$ and used the freedom to insert a contact term $\rho_H\rho_L$ in the second line below) 
\begin{align}
\label{h-idildil}
-\square h_{-i} & = 2\kappa^4\m_H\m_L\Theta(x^+)\[x^+\Theta(x^-)\p_i\(\frac{\p_k \rho_H}{\square_\perp}\p_k\rho_L+\rho_H\rho_L\) - 2\delta(x^-) \frac{\p_i\p_j \rho_L}{\square_\perp} \frac{\p_j \rho_H}{\square_\perp}\]~,\\[5pt]
\label{h--dildil}
-\square h_{--} & = 4\kappa^4\m_H\m_L \delta(x^-) x^+\Theta(x^+) \bigg[ \rho_H\rho_L- \frac{\p_i\p_j \rho_H}{\square_\perp} \frac{\p_i\p_j\rho_L}{\square_\perp}\bigg]~.
\end{align}
Taking the Fourier transform of these expressions (and putting the momenta of the emitted graviton on-shell) gives
\begin{align}
\label{h-idildilFT}
    h_{-i}(k) &= \frac{4\kappa^4\mu_H\mu_L}{k^2+i\epsilon k^-}\int \frac{d^2\bsq_2}{\(2\pi\)^2} \frac{1}{k_+}\[(\bsq_1\cdot \bsq_2) \(q_{2i} - k_i\frac{\bsq_{2}^{2}}{\bsk^{2}}\)-k_i\frac{\bsq_1^2\bsq_2^2}{\bsk^2}\]\frac{\rho_H}{\bsq_1^2}\frac{\rho_L}{\bsq_2^2}~,\\[5pt]
\label{h--dildilFT}
    h_{--}(k) &= \frac{4\kappa^4\mu_H\mu_L}{k^2+i\epsilon k^-}\int \frac{d^2\bsq_2}{\(2\pi\)^2}\frac{1}{k_+^2}\[ (\bsq_1\cdot \bsq_2)^2-\bsq_1^2\bsq_2^2\]\frac{\rho_H}{\bsq_1^2}\frac{\rho_L}{\bsq_2^2}~.
\end{align}
Eqs. \eqref{h-idildilFT} and \eqref{h--dildilFT} are the results quoted in the main text in Eq.~\eqref{hfinalresult}.

\bibliography{apssamp}

\end{document}